\documentclass[conference]{IEEEtran}
\IEEEoverridecommandlockouts
\usepackage{cite}
\usepackage{amsmath,amssymb,amsfonts}
\usepackage{algorithmic}
\usepackage{graphicx}
\usepackage{textcomp}
\usepackage{wrapfig}
\usepackage{caption}
\usepackage{subcaption}
\usepackage{makecell}
\usepackage{url}
\usepackage{listings}
\usepackage{xcolor}
\usepackage{comment}

\definecolor{r1}{RGB}{87,114,158}
\definecolor{r2}{RGB}{204,137,99}
\definecolor{r3}{RGB}{93,157,107}
\definecolor{r4}{RGB}{196,78,82}
\definecolor{r5}{RGB}{129,114,180}
\definecolor{r6}{RGB}{147,120,96}
\usepackage{marginnote}
\usepackage{soul} %
\usepackage{hhline}
\usepackage{ulem}
\definecolor{lightyellow}{RGB}{250, 250, 180}
\definecolor{HLYELLOW}{RGB}{240, 127, 0}
\definecolor{hlyellow}{RGB}{240, 127, 0}
\definecolor{lightcyan}{RGB}{160,255,255}
\sethlcolor{lightcyan}
\usepackage{xparse}

\usepackage{mdframed}
\global\mdfdefinestyle{review}{%
	linecolor=lightcyan,linewidth=3pt,%
	leftmargin=0cm,rightmargin=0cm,%
	skipabove=0cm,skipbelow=0cm,%
	innerrightmargin=0cm,innerleftmargin=0cm,%
	innerbottommargin=0cm,innertopmargin=0cm,%
	backgroundcolor=lightcyan
}
\global\mdfdefinestyle{reviewtext}{%
	linecolor=lightcyan,linewidth=0pt,%
	leftmargin=0cm,rightmargin=0cm,%
	skipabove=0.1cm,skipbelow=0.1cm,%
	innerrightmargin=0cm,innerleftmargin=0cm,%
	innerbottommargin=0cm,innertopmargin=0cm,%
	backgroundcolor=lightcyan
}

\def\BibTeX{{\rm B\kern-.05em{\sc i\kern-.025em b}\kern-.08em
		T\kern-.1667em\lower.7ex\hbox{E}\kern-.125emX}}
\begin{document}
	
	\DeclareDocumentCommand\review{m g g}{%
		{\IfNoValueF {#2}{%
				\IfNoValueF {#3}{%
					{\marginnote{\sethlcolor{#3}\hl{\normalfont \textbf{{\normalsize{\color{white}#2}}}}}%
					}%
				}%
				\IfNoValueT {#3}{%
					{\marginnote{\normalfont \textbf{\normalsize{#2}}}%
					}%
				}%
			}%
			\hl{#1}%
		}%
	}

	\title{HammingMesh: A Network Topology for Large-Scale Deep Learning\\
	}

\author{\IEEEauthorblockN{Torsten Hoefler\IEEEauthorrefmark{1}\IEEEauthorrefmark{2},
		Tommaso Bonato\IEEEauthorrefmark{1}, Daniele De Sensi\IEEEauthorrefmark{1},
		Salvatore Di Girolamo\IEEEauthorrefmark{1}, Shigang Li\IEEEauthorrefmark{1},\\ Marco Heddes\IEEEauthorrefmark{2}, Jon Belk\IEEEauthorrefmark{2}, Deepak Goel\IEEEauthorrefmark{2}, Miguel Castro\IEEEauthorrefmark{2}, and Steve Scott\IEEEauthorrefmark{2}}
	\IEEEauthorblockA{\\
		\IEEEauthorrefmark{1}Department of Computer Science, ETH Zürich, Rämistrasse 101, 8092 Zürich, Switzerland\\
		\IEEEauthorrefmark{2}Microsoft Corporation, One Microsoft Way, Redmond, Washington 98052, United States of America\\
		 \\\IEEEauthorrefmark{1}\{torsten.hoefler, tommaso.bonato, daniele.desensi, salvatore.digirolamo\}@inf.ethz.ch, \{shigangli.cs\}@gmail.com\\
		\IEEEauthorrefmark{2}\{mattheus.heddes, jon.belk, mcastro, steve.scott\}@microsoft.com, \{degoel\}@gmail.com}}

	\maketitle
	
	\begin{abstract}
		Numerous microarchitectural optimizations unlocked tremendous processing power for deep neural networks that in turn fueled the AI revolution.
		With the exhaustion of such optimizations, the growth of modern AI is now gated by the performance of training systems, especially their data movement. 
		Instead of focusing on single accelerators, we investigate data-movement characteristics of large-scale training at full system scale.
		Based on our workload analysis, we design HammingMesh, a novel network topology that provides high bandwidth at low cost with high job scheduling flexibility.
		Specifically, HammingMesh can support full bandwidth and isolation to deep learning training jobs with two dimensions of parallelism. 
		Furthermore, it also supports high global bandwidth for generic traffic.
		Thus, HammingMesh will power future large-scale deep learning systems with extreme bandwidth requirements.
	\end{abstract}
	
	
	\newcommand{\para}[1]{\textcolor{brown}{[#1]}}
	\renewcommand{\para}[1]{}
	\newcommand{\pages}[1]{\textcolor{yellow}{#1 page(s)}}
	\renewcommand{\pages}[1]{}
	\newcommand{\htor}[1]{\textcolor{green}{[htor: #1]}}
	\newcommand{\dan}[1]{\textcolor{blue}{[dan: #1]}}
	\newcommand{\tom}[1]{\textcolor{red}{[tom: #1]}}
	
	\section{Motivation}\pages{1}
	
	\para{dl is important and we should build networks for it!}
	Artificial intelligence is experiencing unprecedented growth providing seemingly open-ended opportunity.
	\emph{Deep learning} models combine many layers of operators into a complex function that is \emph{trained} by optimizing its parameters to large datasets.
	Given the abundance of sensor, simulation, and human artifact data, this new model of designing computer programs, also known as data-driven programming or ``software 2.0''~\cite{sw20}, is mainly limited by the capability of machines to perform the compute- and data-intensive training jobs~\cite{openaicompute}.
	In fact, the predictive quality of models improves as their size and training data grow to unprecedented scales~\cite{kaplan2020scaling}. 
	Building \emph{deep learning supercomputers}, to both explore the limits of artificial intelligence and commoditize it, is becoming not only interesting to big industry but also humanity as a whole.

	\para{specialization opportunities}
	A plethora of different model types exist in deep learning and new major models are developed every two to three years.
	Yet, their computational structure is similar---they consist of layers of operators and they are fundamentally \emph{data-intensive}~\cite{data-movement-is-all-you-need}.
	Many domain-specific accelerators take advantage of peculiarities of deep learning workloads be it matrix multiply units (``tensor cores'')~\cite{tcv100}, specialized vector cores~\cite{xilinxaie}, or specific low-precision datatypes~\cite{msfp}. 
	Those optimizations can lead to orders of magnitude efficiency improvements~\cite{sze,kepner}. 
	Yet, as we are approaching the limits of such microarchitectural improvements, we need to direct our focus to the system level. 
	
	\para{extreme bandwidth is coming and expensive!}
	Today's training jobs are already limited by data movement~\cite{data-movement-is-all-you-need}. 
	In addition, trends in deep neural networks, such as sparsity, further increase those bandwidth demands in the near future~\cite{sparsity-in-dl}.
	Memory and network bandwidth are expensive---in fact, they form the largest cost component in today's systems~\cite{6634083}. 
	Standard HPC systems with the newest InfiniBand adapters can offer 400 Gb/s but modern deep learning training systems offer much higher bandwidths. 
	Google's TPUv2, designed seven years ago, has 1 Tbps off-chip bandwidth~\cite{tpuv2}, AWS' Trainium has up to 1.6 Tbps per Tm1n instance~\cite{trainium}, and Nvidia A100 and H100 chips have 4.8 and 7.2 Tbps (local) NVLINK connectivity, respectively~\cite{a100dgx,h100}.  
	The chips in Tesla's Dojo deep learning supercomputer even have 128 Tbps off-chip bandwidth---\emph{more than a network switch}~\cite{dojo}\setlength{\marginparwidth}{5cm}.
	Connecting these extreme-bandwidth chips at reasonable cost is a daunting task and today's solutions, such as NVLINK, provide only local islands of high bandwidth. 
	
	\begin{figure}[t]
		\begin{center}
			\includegraphics[width=\columnwidth,height=1.2in]{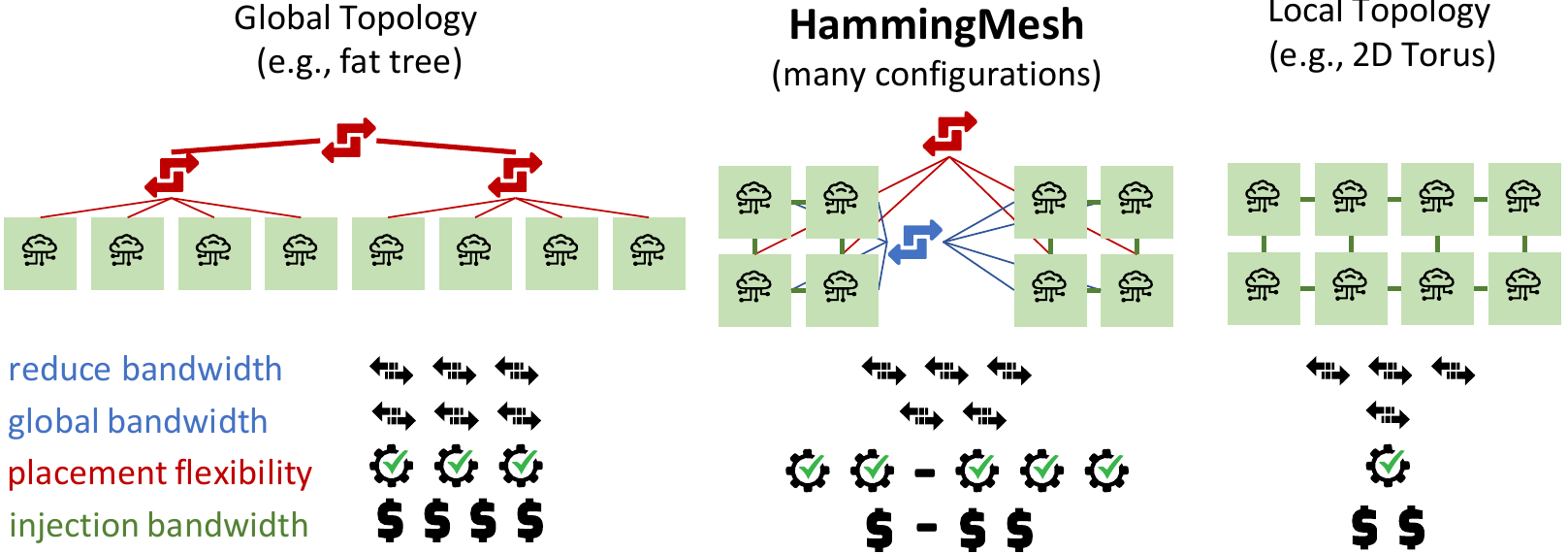}
		\end{center}
		\caption{HammingMesh's bandwidth-cost-flexibility tradeoff.}
		\label{fig:poster}
		\vspace{-1.5em}
	\end{figure}

	\begin{figure*}[t]
		\begin{center}
			\vspace{-1.3em}
			\includegraphics[width=0.99\textwidth,height=1in]{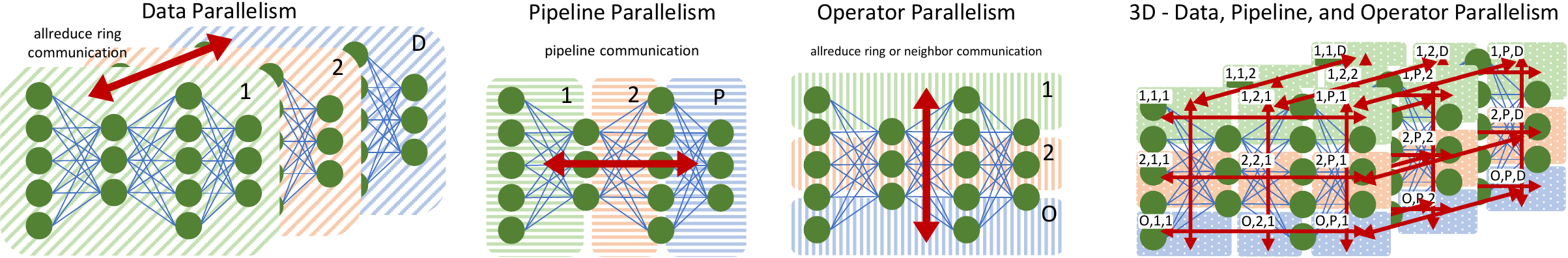}
		\end{center}
		\vspace{-.5em}
		\caption{Distribution strategies for parallel deep neural network training.}
		\label{fig:pardl}
		\vspace{-1.2em}
		
	\end{figure*}

	\para{network specialization}
	We argue that general-purpose HPC and datacenter topologies are not cost-effective at these endpoint injection bandwidths. Yet, workload specialization, similar to existing microarchitectural optimizations, can lead to an efficient design that provides the needed high-bandwidth networking. 
	We begin with developing a generic model that accurately represents the fundamental data movement characteristics of deep learning workloads. 
	Our model shows the inadequacy of the simplistic view that the main communication in deep learning is allreduce~\cite{switchml,sparcml}. In fact, we show that communication can be expressed as a concurrent mixture of pipelines and orthogonal reductions forming toroidal data movement patterns. 
	This formulation shows that today's HPC networks, optimized for full global (bisection) bandwidth, are inefficient for deep learning workloads. Specifically, their \emph{global bandwidth is overprovisioned while their local bandwidth is underprovisioned}. 
	
	We use our insights to develop HammingMesh, a flexible topology that can adjust the ratio of local and global bandwidth for deep learning workloads.
	HammingMesh combines ideas from torus and global-bandwidth topologies (e.g., fat tree) to enable a flexibility-cost tradeoff shown schematically in Figure~\ref{fig:poster}. Inspired by machine learning traffic patterns, HammingMesh connects local high-bandwidth 2D meshes using row and column (blue and red) switches into global networks\footnote{The name \emph{HammingMesh} is inspired by the structural similarity to 2D Hamming Graphs with Meshes as vertices.}.

	\para{principles}
	In summary, we show how deep learning communication can be modeled as sets of orthogonal and parallel Hamiltonian cycles to simplify mapping and reasoning.
	Based on this observation, we define principles for network design for deep learning workloads. Specifically, our HammingMesh topology
	\begin{itemize}
		\item uses technology-optimized local (e.g., PCB board) and global (optical, switched) connectivity.
		\item utilizes limited packet forwarding capabilities in the network endpoints to achieve lower cost and higher flexibility.
		\item enables full-bandwidth embedding of virtual topologies with deep learning traffic characteristics. 
		\item supports flexible job allocation even with failed nodes.
		\item enables flexible configuration of oversubscription factors to adjust global bandwidth.
	\end{itemize}
	With those principles, HammingMesh enables extreme off-chip bandwidths to nearest neighbors at more than 8x cheaper allreduce bandwidth compared to standard HPC topologies such as fat trees.
	HammingMesh reduces the number of external switches and cables and thus reduces overall system cost. Furthermore, it provides significantly higher flexibility than torus networks. 
	HammingMesh also enables seamless scaling to larger domains without separation between on- and off-chassis programming models (like NVLINK vs. InfiniBand). 
	And last but not least, we believe that HammingMesh topologies extend to other machine learning, (multi)linear algebra, parallel solvers, and many other workloads with similar traffic characteristics.

	We start with a characterization of parallel deep learning and the related data movement patterns. 
	For reference, we provide an overview of symbols used in this paper in Table~\ref{symbols}.
	\begin{table}[h!]
			\begin{center}
				\begin{small}
					\begin{tabular}{c | c} 
						Symbol & Description \\ [0.1ex] 
						\hline %
						$M$ & number of examples per minibatch \\ 
						$N_\mathcal{P}$ & number of network parameters\\
						$W$ & size of a word\\\hline
						$D$, $P$, $O$ & degree of data, pipeline, operator parallelism\\\hline
						$a$, $b$ and $x$, $y$ & 2D HammingMesh board and global sizes\\
					\end{tabular}
				\end{small}
			\end{center}
			\caption{Symbols used in the paper}
			\vspace{-2em}
			\label{symbols}		
	\end{table}

	\section{Communication in Distributed Deep Learning}\pages{2}
	\label{sec:comm}
	One iteration of deep learning training with Stochastic Gradient Descent (SGD) consists of two phases: the forward
	\begin{wrapfigure}[5]{r}{1.3in}
			\begin{center}
				\vspace{-1.5em}
				\includegraphics[width=1.3in]{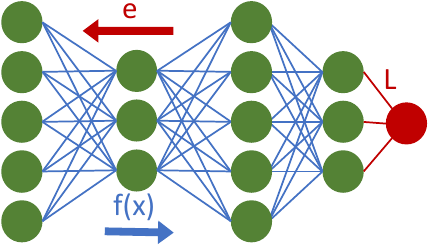}
			\end{center}
	\end{wrapfigure}
	pass and the backward pass.
	The forward pass evaluates the network function $f(x)$ on a set of $M$ examples, also called a  ``minibatch''.
	The backward pass of SGD computes the average loss $L$ and propagates the errors $e$ backwards through the network to adapt the parameters $\mathcal{P}$. 
	This training process proceeds through multiple (computationally identical) iterations until the model achieves the desired accuracy.
	
	Parallelism and data distribution can fundamentally be arranged along three axes: \emph{data parallelism}, \emph{pipeline parallelism}, and \emph{operator parallelism}~\cite{distdl-preprint}. The latter two are often summarized as \emph{model parallelism} and operator parallelism is sometimes called tensor parallelism~\cite{trainium}. We now briefly discuss their main characteristics.
	
	\subsection{Data parallelism}
	
	When parallelizing over the training data, we train $D$ separate copies of the model, each with different examples. To achieve exactly the same result as in serial training, we sum the distributed gradients before applying them to the weights at the end of each iteration. If the network has $N_\mathcal{P}$ parameters, then the communication volume of this step is $WN_\mathcal{P}$.

	Modern deep neural networks have millions or billions of parameters, making this communication step expensive. Thus, many optimizations target gradient summation~\cite{sparcml}---some even change convergence properties during the training process but maintain final result quality~\cite{top-k}. 
	Dozens of different techniques have been developed to optimize this communication---however, all perform some form of distributed summation operation like \textsl{MPI\_Allreduce}.
	Data-parallelism differs thus mostly in the details such as invocation frequency, consistency, and sparsity~\cite{localsgd1,stich2019local,gorbunov2020local,khaled2020tighter,dryden-quant,sparcml,NEURIPS2018_3328bdf9,NEURIPS2018_31445061,45187}.
	We describe various schemes in Appendix~\ref{app:datapar} in more detail.

	\begin{figure*}[t]
		\begin{center}
			\vspace{-1.3em}
			\includegraphics[width=0.9\textwidth]{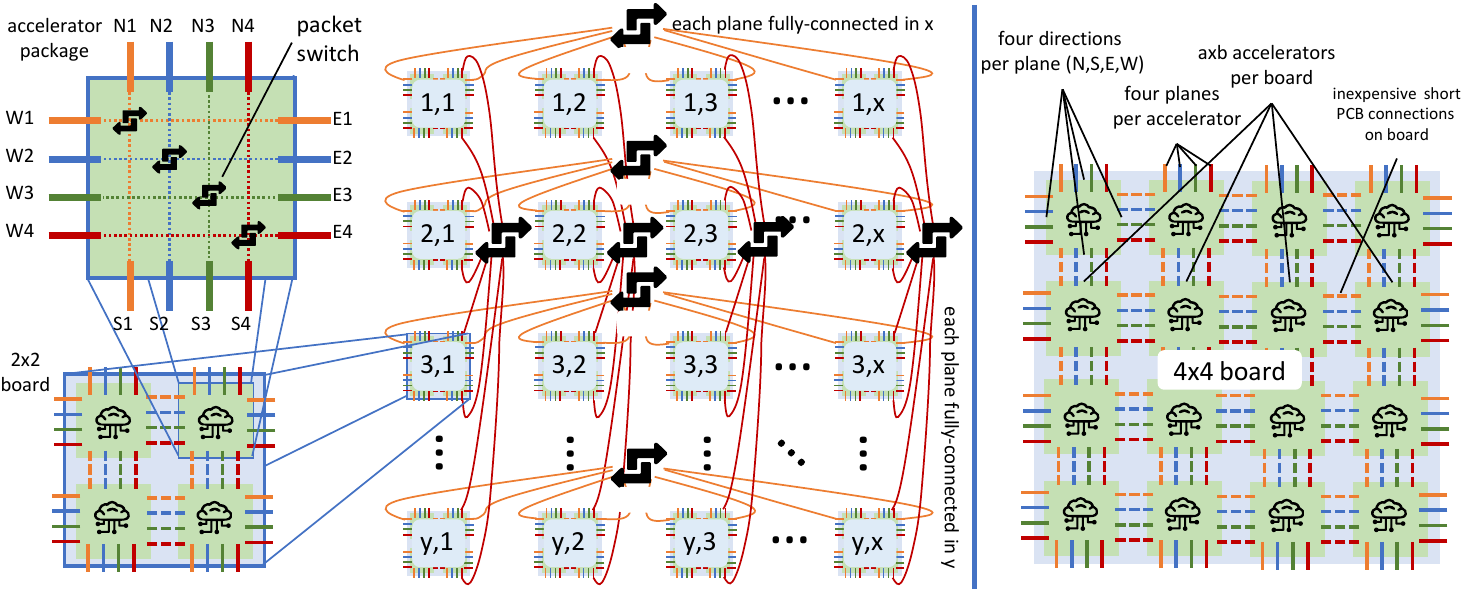}
		\end{center}
		\vspace{-.5em}
		\caption{HammingMesh structure: left $x\times y$ Hx2Mesh, right Hx4Mesh board, both with four planes.}
		\label{fig:hxmesh}
		\vspace{-1.2em}
	\end{figure*}
	
	\subsection{Pipeline parallelism}
	
	Deep neural networks are evaluated layer by layer with the outputs of layer $i$ feeding as inputs into layer $i+1$. Back-propagation is performed along the reverse direction starting at the loss function $L$ after the last layer and proceeding from layer $i+1$ to layer $i$. 
	We can model the network as a pipeline with $P$ stages with one or more layers per stage~\cite{huang2019gpipe}. 
	Forward and backward passes can be interleaved at each processing element to form a bidirectional training pipeline~\cite{harlap2018pipedream}.
	Pipelines suffer from characteristic start-up and tear-down overheads. These can be reduced by running two pipelines in both directions~\cite{chimera} or by using asynchronous schemes that impact convergence~\cite{MLSYS2021_6c8349cc}. 
	
	Overall, pipelining schemes can use $P$ processors with a nearest-neighbor communication volume proportional to the number of output activations at the cut layers.

	\subsection{Operator parallelism}
	
	Very large layer computations (operators) can be distributed to $O$ processors. Most deep learning layer operators follow computational schedules of (multi-)linear algebra and tensor contractions and require either (tightly-coupled) distributed reductions or nearest-neighbor communications. 
	We discuss communication schemes of typical operators in Appendix~\ref{app:oppar}.

	\subsection{Overall communication pattern}\label{sec:dlcomm}
	When all forms of parallelism are used, then the resulting job comprises $D\times P\times O$ accelerators; each accelerator in a job has a logical address $(1..D,1..P,1..O)$. The data-, pipeline-, and operator-parallel communication can be arranged as one-dimensional slices (rings) by varying only one coordinate of the Cartesian structure. Pipelines would leave one connection of the ring unused.
	For example, the data-parallel dimension consists of $P\cdot O$ rings of length $D$ each. Each of those rings represents a single allreduce.
	We show efficient ring-based reduction and broadcast algorithms for large data volumes in Section~\ref{sec:fastred}.
	
	The overall composition of communication patterns forms a torus as illustrated in the right part of Figure~\ref{fig:pardl} for a 3$\times$3$\times$3 example: Both the operator and the data parallel dimensions use nine simultaneous allreductions of size three each. The pipeline parallel dimension uses nine three-deep pipelines on three different model replicas, each split in three pieces. 
	
	While we can map such a logical torus to a full-bandwidth network topology, it seems wasteful to provide full bandwidth for sparse communication. For example, a $400$ Gb/s nonblocking fat tree with 16,384 endpoints provides full bisection bandwidth of more than $\frac{16,384\cdot \mathrm{50\,GB/s}}{2}=410$ TB/s. A \reversemarginpar bi-directional \normalmarginpar 32x32x16 torus communication pattern requires at most $32\cdot 16\cdot 2 \cdot 50$ GB/s$=51.2$ TB/s bisections (cutting one dimension of size 32) - a mere 12.5\% of the offered bandwidth. 
	In other words, \emph{88\% of the available bandwidth will remain unused and is wasted}. 
	Furthermore, it is not always simple to map such torus communication patterns efficiently to full-bandwidth low-diameter topologies in practice~\cite{prisacari-dragonfly-mapping}.

	\begin{table*}
		\begin{center}
			\begin{small}
				\begin{tabular}{c | c c c c c c | c c c c c c }  
					&     \multicolumn{6}{c}{Small Cluster ($\approx$1,000 accelerators)} &  \multicolumn{6}{c}{Large Cluster ($\approx$16,000 accelerators)} \\
					Topology &  \makecell{cost \\ $[$M\$$]$} & \makecell{glob. BW\\ $[$\% inject$]$} & \makecell{global\\saving} & \makecell{ared. BW\\ $[$\% peak$]$} & \makecell{ared.\\saving} & diam. & \makecell{cost \\ $[$M\$$]$} & \makecell{glob. BW\\ $[$\% inject$]$} & \makecell{global\\saving} & \makecell{ared. BW\\ $[$\% peak$]$} & \makecell{ared.\\saving} & diam. \\ [0.1ex] 
					\hline %
					
					nonbl. FT    & 25.3 & 99.9 & 1.0x & 98.9 & 1.0x & 4 &  680 & 98.9  & 1.0x & 99.8 & 1.0x & 6\\ 
					50\% tap. FT & 17.6 & 51.2 & 0.7x & 98.9 & 1.4x & 4 &  419 & 47.6  & 0.8x & 99.8 & 1.6x & 6\\
					75\% tap. FT & 13.2 & 25.7 & 0.5x & 98.9 & 1.9x & 4 &  271 & 24.0  & 0.6x & 99.8 & 2.5x & 6\\
					
					\hline
					
					Dragonfly & 27.9 & 62.9 & 0.6x & 98.8 & 0.9x & 3 &  429 & 71.5  & 1.2x & 98.6  & 1.6x & 5\\
					
					\hline
					
					2D HyperX\footnotemark & 10.8 & 91.6 & \textbf{2.1x} & 98.1 & 2.3x & 4 &  448 & 95.8  & 1.5x & 91.4  & 1.4x & 8\\
					
					\hline
					
					Hx2Mesh & 5.4  & 25.4 & 1.2x          & 98.3 & 4.7x          & 4 & 224  & 25.0 & 0.8x & 92.3 & 2.8x & 8\\
					Hx4Mesh & 2.7  & 11.3 & 1.0x & 98.4 & \textbf{9.3x} & 8 & 43.3 & 10.5 & \textbf{1.7x} & 92.2 &  \textbf{14.5x} & 8\\
					
					\hline
					
					2D torus & 2.5 & 2.0 & 0.2x & 98.1 & 10.1x & 32 & 39.5 & 1.1  & 0.2x & 91.4  & 15.7x & 128\\	
				\end{tabular}
			\end{small}
		\end{center}
		\vspace{-.9em}
		\caption{Overview of our example networks (small and large cluster) using the cost model in Section~\ref{sec:cost}. All bandwidths are the result of the packet-level simulations detailed in Section~\ref{sec:microbench}. Global alltoall bandwidth is reported as share of the injection bandwidth for large messages (1.6 Tb/s). Allreduce bandwidth is reported as share of the theoretical optimum (1/2 of the injection bandwidth) for large messages. The cost savings for global and allreduce bandwidth are relative to the corresponding network cost of the nonblocking fat tree. Note that the diameter counts all cables and its derivation is explained in Section~\ref{sec:diam}.}
		\label{tab:overview}
		\vspace{-1.5em}
	\end{table*}

	\section{HammingMesh}\pages{2}
	
	Based on the communication workload analysis, we now design a flexible and efficient network topology. The basic requirements are to support highest injection bandwidth for a set of jobs, each following a virtual toroidal communication topology. 
	We note that medium-size models are often decomposed only in two dimensions in practice (usually data and pipeline or data and operator).
	Only extreme-scale workloads require all three dimensions---even then, communication along the data parallel dimension only happens after one complete iteration. Thus, we use a two-dimensional physical topology.

	As a case study, we assume a modern deep learning accelerator package with 16 400 Gb/s off-chip network links, a total network injection bandwidth of 800 GB/s (top left in Figure~\ref{fig:hxmesh}). 
	Our topology design also takes technology costs into account: Similar to Dragonfly, which combines local short copper cables with global long fiber cables to design a cost-effective overall topology, we combine such local groups with a global topology.
	Different from Dragonfly, we choose two quite distinct topologies: The local groups are formed by a local inexpensive high-bandwidth 2D mesh using short metal traces on PCB boards. This is the opposite of  Dragonfly designs, which combine densely-connected local groups (``virtual switches'') and connect those fully globally. HammingMesh combines sparsely connected boards in a dimension-wise (not globally)  fully-connected topology.
	Those boards are connected by a two-dimensional Hamming graph, in which each dimension is logically fully connected (e.g., by a fat tree).
	All accelerator ports are arranged in {\em planes} with four directions each. Our example accelerator has four planes (top left in Figure~\ref{fig:hxmesh}), e.g.,  plane 1 has ports E1, W1, N1, and S1.
	We assume that each accelerator can forward packets within a plane like any network switch. Accelerators do not have to forward packets between planes, e.g., packets arriving at N1 may only be forwarded to E1, W1, or S1 but none of the other ports. Thus, only simple 4x4 switches are needed at each accelerator. Figure~\ref{fig:hxmesh} illustrates the structure in detail.
	
	A 2D HammingMesh is parameterized by its number of planes and four additional numbers: $(a,b)$, the dimensions of the board, and $(x,y)$, the dimensions of the global topology.
	It connects a total of $abxy$ accelerators. 
	We abbreviate HammingMesh with HxMesh in the following. Furthermore, an HxMesh with an $a\times b$ accelerator board is called H$a$x$b$Mesh, e.g., for a 2x2 board, H2x2Mesh. For square board topologies, we skip the first number, e.g., an H2x2Mesh that connects 10x10 boards is called a 10x10 Hx2Mesh. 
	
	HxMesh has a large design space: We can combine different board and global topologies, e.g., 3D mesh boards with global Slim Fly topologies~\cite{SF}. In this work, we consider 2D boards as most practical for PCB traces. The board arrangement could be reduced to a 1D HxMesh, where $y=1$ and each N$k$ link is connected to the corresponding S$k$ link (``wrapped around''). The same global topology can also span multiple rows or columns (e.g., full boards in a single fat tree).
	For ease of exposition, we limit ourselves to 2D HxMeshes using 2D boards and row/column-separated global topologies.  
	We use two-level fat trees as global topologies to connect the boards column and row wise. If the boards can be connected with a single 64-port switch, we use that instead of a fat tree.
	
    \footnotetext{Note that a 2D HyperX is identical to an Hx1Mesh}
	\subsection{Bisection and global bandwidth}\label{sec:bisection}
	
	Bisection cut is defined as the minimal number of connections that would need to be cut in order to bisect the network into two pieces, each with an equal number of accelerators. The bisection bandwidth is the cut multiplied by the link bandwidth. 
	Let us assume a single-plane of an $x\times y$ Hx$a$Mesh (square board) with $x\leq y$ and $y$ even, wlog. We now consider the $xy/2$ ``lower'' half boards with $y$ coordinates $1,2,\ldots y/2$. We split the HxMesh into two equal pieces by cutting the $2a$ links in $y$ direction of each of the lower half of the boards. This results in a total cut width of $axy$. 
	Each accelerator has four network links per plane, a total injection bandwidth of $4a^2$ per board. 
	We have $xy/2$ boards with a total injection bandwidth of $4a^2xy/2=2xya^2$ in each partition.
	Thus, the relative bisection bandwidth is $axy/2xya^2=1/2a$.

	\para{globall bandwidth can be higher than bisection bandwidth}
	In a bisection traffic pattern, all traffic crosses the network bisection (any two communicating endpoints are in different sets of the bisection). Such (worst-case) patterns are rare in practice. 
	A more useful pattern, more often observed in practice is alltoall, where each process sends to all other processes. This pattern is the basis of parallel transpositions, Fast Fourier Transforms, and many graph algorithms.
	The achievable theoretical bandwidth for such alltoall patterns is often called ``global bandwidth''.
	Some topology constructions take advantage of the fact that global bandwidth is higher than bisection bandwidth. Prisacari et al.~\cite{prisacari-ics-bandwidth-opt-alltoall} shows that full-global bandwidth (alltoall) fat trees can be constructed with 25\% less switches than nonblocking fat trees. Dragonfly~\cite{DF}, Slim Fly~\cite{SF}, or other low-diameter topologies~\cite{diameter-2-topos} can further reduce the number of switches in very large installations while maintaining full global bandwidth.
	As is customary for low-diameter topologies~\cite{DF,SF}, we assess it using packet-level simulations of alltoall traffic.

	\subsection{Network diameter}\label{sec:diam}
	
	We now analyze the diameter of the different topologies by counting cables between source and destination. For example, a two-level fat tree has diameter four, while many works discount the cable to/from the endpoint, we count it to ensure fairness with direct topologies such as simple torus networks.
	
	To compute the diameter of HxMesh, we consider the distance between two accelerators on two different boards. Both accelerators are as distant as possible from the edges of the corresponding boards, and source and destination boards are on different rows and columns. $\lfloor (a-1)/2\rfloor$ hops lead from an inner accelerator on a board to the East or West edge. The switch allows to take two cables to an intermediate board (sharing the same HxMesh row of the source board, and the same HxMesh column of the destination board). Once on the intermediate board, we need another $\lfloor (a-1)/2\rfloor$ hops to reach the correct board column. From there, we need $\lfloor (b-1)/2\rfloor$ hops to reach the North or South edge, and then two cables (through the switch) to reach the destination board. Eventually, other $\lfloor (b-1)/2\rfloor$ hops are needed on the destination board to reach the destination accelerator.
	
	The previous discussion assumes a flat (alltoall) global topology, which is not practical for large $x$ and $y$. If we instead use a full bandwidth fat tree built from routers with $k$ ports for global connectivity, it has a diameter of $2(\lceil \log_{k/2} q/k\rceil+1)$, where $q$ is the number of endpoints. Because in HxMesh a tree connects either boards on the same column or on the same row, and because each tree is connected to two opposite edges of a board, we either have $q=2x$ or $q=2y$.
		Thus, the HxMesh diameter is $2(\lfloor (a-1)/2\rfloor + \lfloor (b-1)/2\rfloor) + 2(\lceil \log_{k/2} 2x/k\rceil+1) +$\\$+ 2(\lceil \log_{k/2} 2y/k\rceil+1)$.
	
	\subsection{Cost model}\label{sec:cost}
	
	We analyze capital expenditure when purchasing the system in the following. We assume that the accelerator NICs and ports as well as the PCB are included in the endpoint packaging cost for all topologies. 
	We charge for the network equipment: optical transceivers, cables, and switches. 
	We use three types of cables and a single type of switch with costs from Colfaxdirect (in April 2022). A single 64-port switch costs \$14,280. One 20m Adaptive optical Cable (AoC) costs \$603 and one 5m Direct Attach Cables (DAC) costs \$272. All topologies can be built with these two types of cables using standard 19 inch racks. To compute the costs in Table~\ref{tab:overview}, we use DAC cables to connect to the endpoints and AoC cables between switches.
	More details are found in Appendix~\ref{app:pricing}.

	\subsection{Example topologies}
	
	We consider a small cluster with approximately 1,000 accelerators and a large cluster with approximately 16,000 accelerators as specific design points to compare realistic networks. 
	We compare various fat trees  (nonblocking, 50\%, 75\% tapered), full bandwidth Dragonfly, and two-dimensional torus, with Hx2Mesh and Hx4Mesh example topologies. 
	
	Table~\ref{tab:overview} summarizes the main cost and bandwidth results. Global and allreduce bandwidth are determined using packet-level simulations (see Section~\ref{sec:res}) for large messages. \emph{For all experiments, we simulated a single plane of HammingMesh and four planes for all other topologies, i.e., a total injection bandwidth of 4$\times$400 Gb/s.}
	We use industry-standard layouts and cable configurations for the cost estimates: 
	fat trees are tapered beginning from the second level and connect all endpoints using DAC and all switches using AoC. 
	Dragonfly topologies use full-bandwidth groups with $a=16$ routers each, $p=8$ endpoints per router, and $h=8$ links to other groups with DAC links inside the groups and AoC links between groups. 
	The torus uses $2\times 2$ board topologies with discounted local PCB connectivity, similar to Hx2Mesh and only DAC cables between the boards. 
	For HxMeshes, we use DAC links to connect endpoints to switches along one dimension, and AoC links for the other dimension. All inter-switch links are AoC as in fat trees.
	All details and exact cable counts are described in Appendix~\ref{app:topos}.

	\subsection{Logical job topologies and failures in HxMesh}
	
	\para{logical toplogies}
	As we discussed in Section~\ref{sec:dlcomm}, communication patterns in deep learning can be modeled as sets of cycles. 
	Typical learning jobs use either logical 1D cycles for small models with only data parallelism or 2D tori that combine data and pipeline parallelism for medium-scale models or combining pipeline and model parallelism for very large models. 
	Each specific training job will have a different optimal decomposition resulting in 1D, 2D, or sometimes even 3D logical communication topologies. 
	
\begin{figure}[t]

			\begin{center}
				\includegraphics[width=\columnwidth]{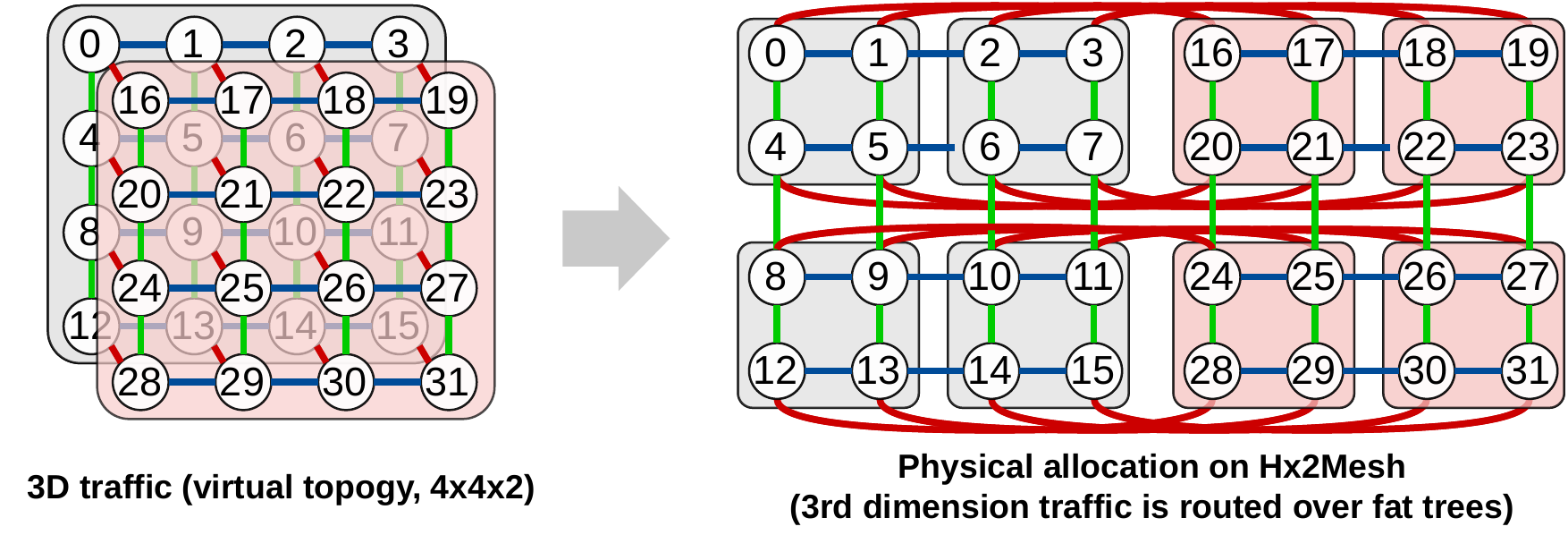}
			\end{center}
			\caption{3D workload mapping onto Hx2Mesh example. Left: virtual 4x4x2 topology. Right: mapping on Hx2Mesh.}
			\vspace{-1.7em}
			\label{fig:3dtraffic}
	\end{figure}

	\para{definition of training job}
	We use logical 2D topologies for our training jobs. Each job uses several boards and requests a $u\times v$ layout (i.e., $a$, $b$ divides $u$, $v$, respectively). 
	If the application topology follows a 1D or 3D scheme, then users use standard folding techniques to embed it into two dimensional jobs~\cite{arrays-trees-hypercubes}. 
	Figure~\ref{fig:3dtraffic} shows an example of 3D virtual topology mapped on an Hx2Mesh physical topology. 
		Processes can be sliced on the third dimension and mapped on different boards. 
		Communications between different slices of the third dimension are routed over the per-column or per-row fat trees, depending how different slices are mapped. 
		To minimize communication latency between slices, consecutive slices should be adjacent to each other.
	Furthermore, we will show in Section~\ref{sec:fastred} how to accelerate allreduce on a 2D torus instead of a ring.

	\para{logical subnets}
	It is easy to see that any consecutive $u\times v$ block of boards in a 2D HxMesh has the same properties as a full $u\times v$ HxMesh. We call such subnetworks \emph{virtual sub-HxMeshes}. They are a major strength of HxMesh compared to torus networks in terms of fault tolerance as well as for allocating jobs. 
	In fact, HxMeshes major strength compared to torus networks is that virtual subnetworks can be formed with non-consecutive sets of boards (not only blocks): any set of boards in an HxMesh where all boards that are in the same row have the same sequence of column coordinates can form a virtual subnetwork. We will show examples below together with a motivation for subnetworks---faults.
	
	\paragraph{Fault-tolerance}
	
	We assume that a board is the unit of failure in an HxMesh, i.e., if an accelerator or link in a board fail, the whole board is considered failed. This simplifies system design and service. Partial failure modes (e.g., per plane) are outside the scope of this work.
	
\begin{figure}[h!]
		\begin{center}
			\vspace{-1.2em}
			\includegraphics[width=\columnwidth]{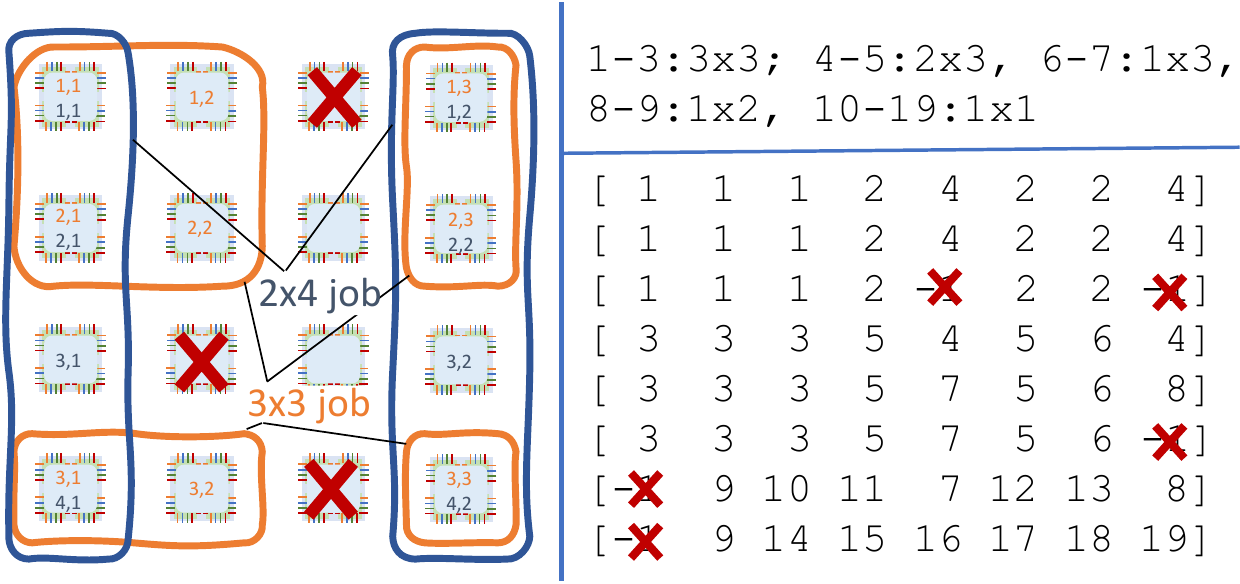}
		\end{center}
		\vspace{-1em}
		\caption{Subnetworks in the case of failures}
		\vspace{-1.5em}
		\label{fig:subnets}
	\end{figure}

	The left part of Figure~\ref{fig:subnets} shows a 4x4 Hx2Mesh and three board failures. 
	We show two different subnetworks (many more are possible): a 2x4 subnetwork (blue) with the physical boards $(1,1), (1,4), (2,1), (2,4), (3,1), (3,4), (4,1), (4,4)$  and a 3x3 subnetwork (yellow) with the physical boards $(1,1), (1,2), (1,4), (2,1), (2,2), (2,4), (4,1), (4,2), (4,4)$. We also annotate the new coordinates of boards in the virtual subnetworks. Remapping can be performed transparently to the user application, which does not observe a difference between a virtual and physical HxMesh in terms of network performance. 
	The right part of the figure shows the output of our automatic mapping tool for a more complex configuration of jobs (top, read job ids 1-3 are $3\times 3$ logical jobs etc.). We analyze in detail the effects of fragmentation due to failed boards in Section~\ref{sec:expwrkl}
	
	\subsection{Tapering the dimensions}
	
	As we discussed in Section~\ref{sec:dlcomm}, deep learning workloads do not require much global bandwidth.
	We can thus reduce the global bandwidth of HxMeshes further by tapering each dimension to reduce the cost.
	
	First, we observe that for mapping a single ring along one dimension of an HxMesh, we only need two ports between neighboring switches in that direction.
	Consider the Hx2Mesh in Figure~\ref{fig:hx2silly} with a two-level fat tree to implement the global connections and a ring spanning all accelerators in x direction (we show only one of the 8 identical physical rings for clarity):
	\begin{figure}[h!]

			\begin{center}
				\vspace{-1.7em}
				\includegraphics[width=\columnwidth, height=1in]{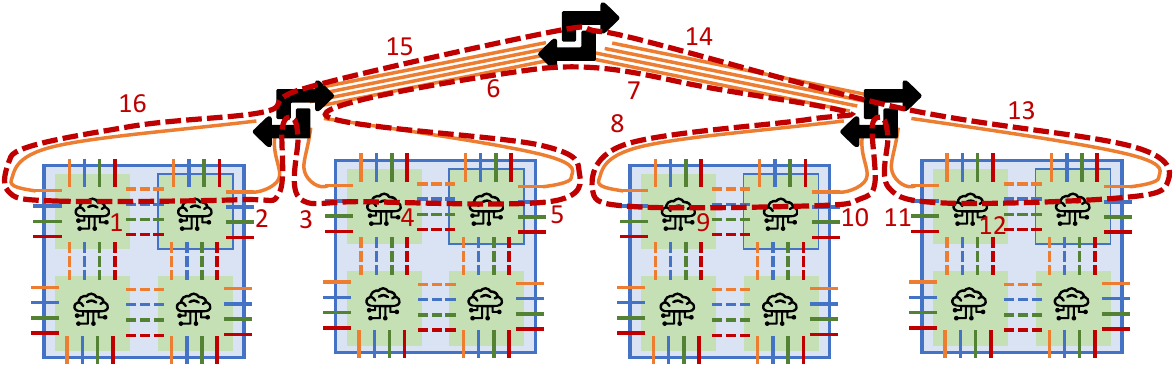}
			\end{center}
			\caption{Ring mapping example in a two-level fat tree.}
			\vspace{-.9em}
			\label{fig:hx2silly}
	\end{figure}
	
	We observe that switches in the top level only route two connections between neighboring switches. In this example, only 50\% of the links are utilized. 
	More generally, we rarely need full bandwidth in the global topology. We only need it for global traffic and to support fragmented (non-consecutive) allocations. 
	For example, for a large deep learning system, we may only need 20\% global bandwidth to support global traffic and 
	moderate fragmentation. This provides a second dial in addition to the board size to adjust the tradeoff between global bandwidth and cost.

	\section{Using HammingMesh in Practice}
	
	Training jobs request two-dimensional sets of boards. The allocation of rectangular jobs on a 2D torus is equivalent to the strongly NP-hard 2D bin packing problem~\cite{gareyjohnson}. Yet, HxMesh allocations support splitting blocks as illustrated in Figure~\ref{fig:subnets}, significantly simplifying the problem. We now show a simple and effective greedy allocation strategy.
	
	\subsection{Allocating jobs on HammingMesh}\label{sec:alloc}
	We develop the following simple greedy allocation strategy for an $au\times bv$ job to an $x\times y$ HxMesh:
	
	\begin{enumerate}
		\item Identify all available indexes in each row, resulting in $y$ sets of at most $x$ indexes.
		\item Set $\mathcal{S}$ (``selected'') to the first row with at least $v$ indexes.
		\item Add another row whose intersection with all rows in $\mathcal{S}$ has at least $v$ indexes to $\mathcal{S}$. %
		\item Repeat the last step until $\mathcal{S}$ contains $u$ rows, fail if no such set exists.
	\end{enumerate}
	
	We implemented this procedure in less than 50 lines of Python and allocated an extremely large 1,000x1,000 HxMesh in less than one second on a laptop. We now outline a set of simple optimization heuristics that can be used to improve utilization of HxMeshes.

	\textbf{Transpose} If it fails to find an $u\times v$ block, then we retry to find a (transposed) $v\times u$ block. 
	
	\textbf{Aspect ratio} Jobs could also allow to change their aspect ratio. For example, a job requesting 4x16 boards may also function well with 2x32 boards. 
	
	\textbf{Sorting} If the jobs to be allocated are known in advance, they can be allocated from the largest to the smallest to reduce fragmentation.  
	
	\textbf{Locality} The allocation algorithm can evaluate different aspect ratios and select the one that, for alltoall traffic, minimizes the traffic in the upper levels of the trees connecting the boards (so that performance is less affected by tapering).
	
	We analyze the impact of these optimization on the system utilization below.

	\paragraph{Job interference}
	A valuable property of this allocation scheme is that it avoids network interference between different jobs. 
	Because each board is used at most by one job at a time, interference can never occur within the board. 
	Even when source and destination are on two different boards on the same row (or on the same column) a nonblocking fat tree with packet spraying can avoid interference. 
	When source and destination boards are on different rows and different columns, packets will traverse an intermediate board (Section~\ref{sec:routing}). 
	Our allocation scheme ensures that each board shares at least the row or the column with another board belonging to the same job (e.g., see Figure~\ref{fig:subnets}). Thus, packets never cross boards belonging to a different job to reach any destination.
	
	\paragraph{Defragmentation}
	Modern cluster and cloud systems supports efficient checkpoint/restart. Thus, we assume that we can \emph{defragment} the system by checkpointing jobs, shuffling them, and restarting them in a better permutation. 
	Our example accelerator can send 64 GiB in less than 80 ms. Thus, a system with reasonable global bandwidth (e.g., 10\%) can be defragmented in less than a second. Given the long running times of deep learning workloads (hours to days), we expect that such defragmentation will not impact the user experience or utilization of the overall system during operation.
	
	\subsection{Experimental workloads}\label{sec:expwrkl}
	
\begin{figure}[h!]
		\begin{center}
			\includegraphics[width=\columnwidth]{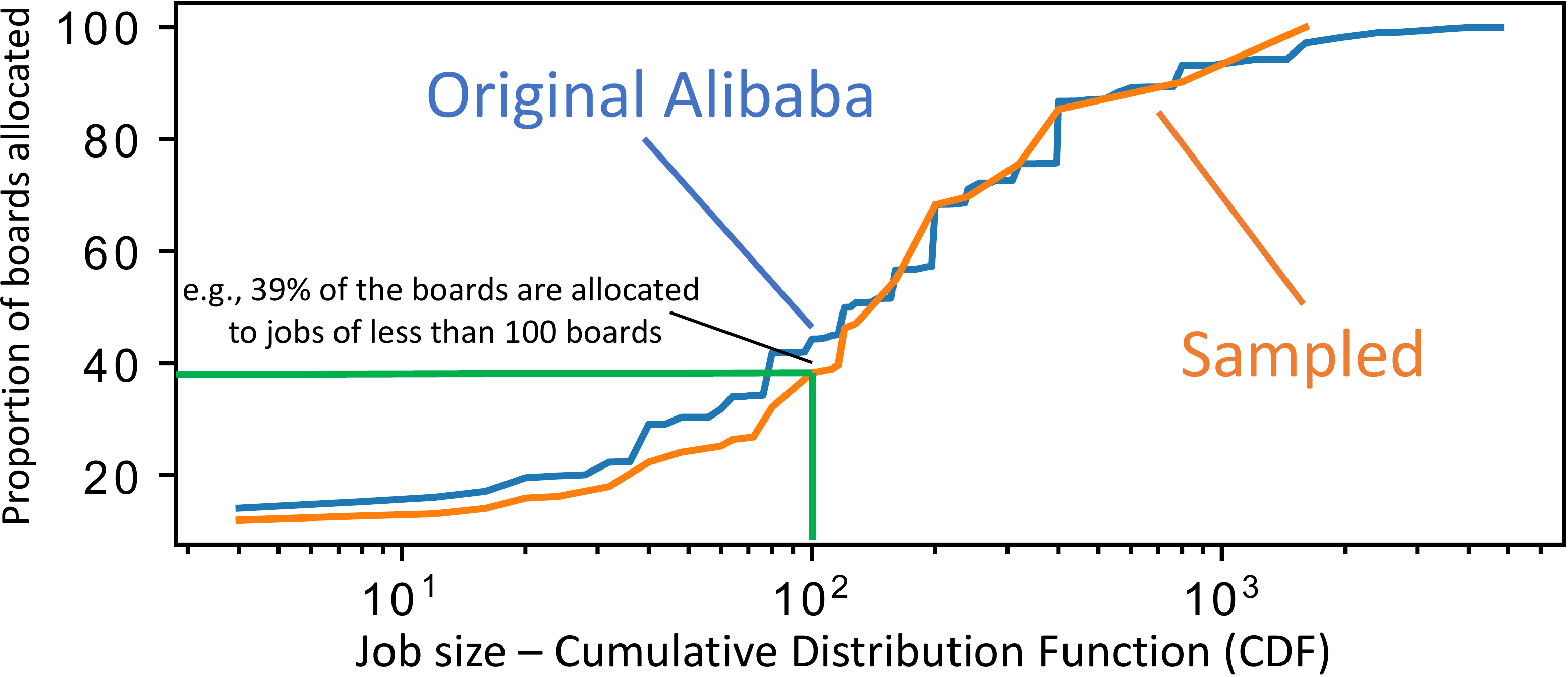}
		\end{center}
		\caption{Proportion of boards allocated to a certain size job.}
		\label{fig:jobs-cdf}
		\vspace{-0.5em}
	\end{figure}

	We analyze the quality of our allocation algorithm using the distribution of job sizes extracted from a two-month workload log of Alibaba's ML-as-a-service (MLaaS) cluster with 6,742 GPUs~\cite{276938,alibaba-trace}. 
	We aim to simulate how well a representative job mix that completely fills a full global bandwidth topology can be allocated on an HxMesh. For this, we draw 1,000 such job mixes from the job size distribution. We do this by sampling a job size, multiply it by the size of the board, and add it to the cluster. We repeat this step until we fill the target cluster fully (we carry samples that do not fit to the next jobs mix). We store the (random) order of drawn samples in a job trace. 
	Figure~\ref{fig:jobs-cdf} shows the cumulative distribution function of the proportion of boards allocated to jobs of a specific size for both the original distribution and the sampled distribution that fully occupy the Alibaba cluster. 
	
\begin{figure}[h!]
		\begin{center}		
			\includegraphics[width=\columnwidth]{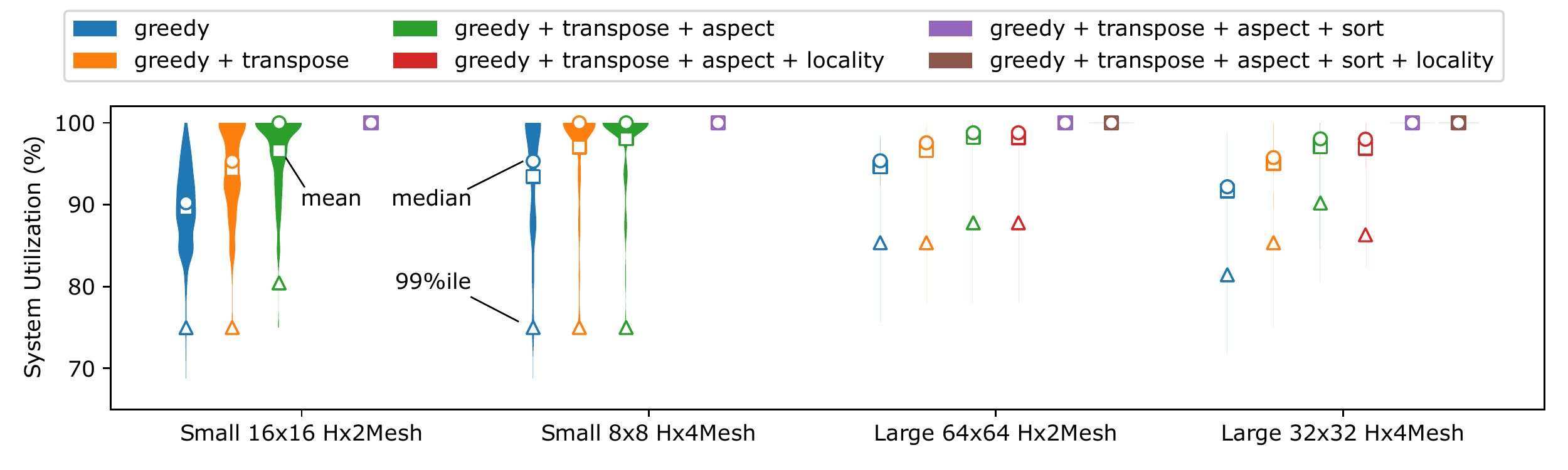}
		\end{center}
		\caption{System utilization using different optimizations. Each color corresponds to a specific set of optimizations.}
		\label{fig:alloc-improvements}
		\vspace{-0.5em}
	\end{figure}

	Figure~\ref{fig:alloc-improvements} shows the system utilization of our greedy allocation algorithm and different optimization heuristics on the different HxMeshes described in Table~\ref{tab:overview}. The figure shows the distribution of 1,000 allocations of random job traces. Full-global bandwidth topologies achieve 100\%. 
	By default, we make jobs as square as possible. The aspect ratio optimization allows changing the aspect ratio up to eight. 
	
	We observe that, even without any optimization, the greedy algorithm leads to a 90\% system utilization. When transposing the jobs this further increases by an additional 5-8\%. If jobs are further sorted by their size, we observe a mean and median utilization higher than 98\%, with the 99th percentile higher than 95\%.
	
\begin{figure}[h!]
		\begin{center}
			\includegraphics[width=\columnwidth]{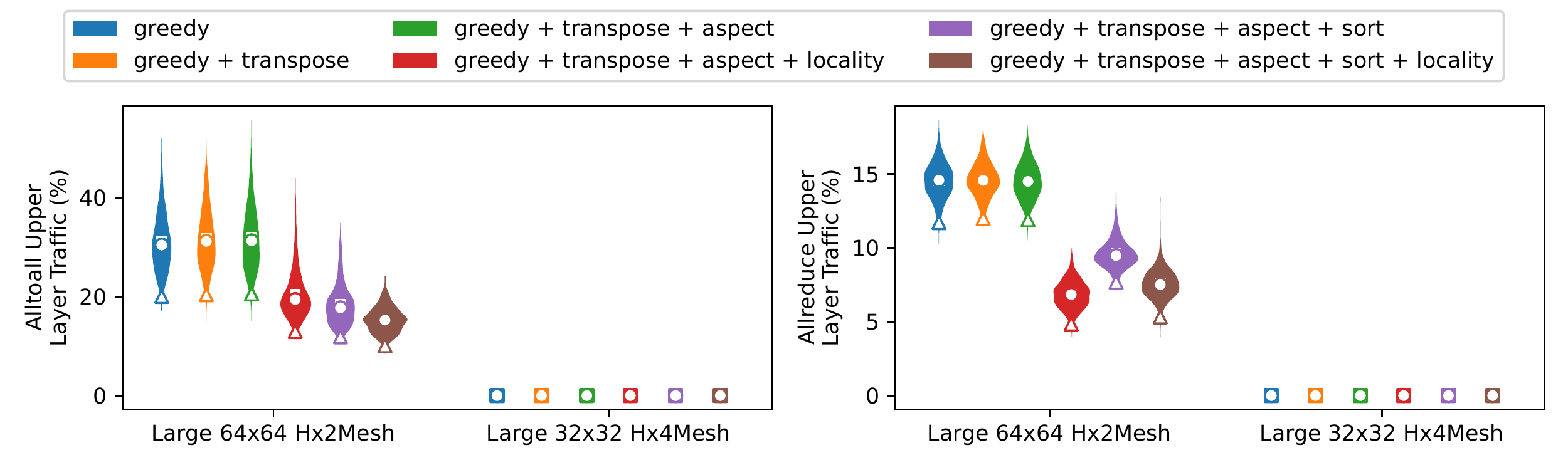}
		\end{center}
		\caption{Fraction of traffic crossing upper fat tree levels.}
		\vspace{-.9em}
		\label{fig:alloc-improvements-links}
		\vspace{-0.5em}
	\end{figure}

	We now investigate the expected load on the upper levels of the fat tree in the large cluster variants. 
	Figure~\ref{fig:alloc-improvements-links} shows the fraction of traffic crossing the upper levels of the fat trees, both for jobs running alltoall traffic or allreduce traffic, respectively. 
	
	We observe that in both cases the percentage of traffic crossing the upper level is lower than 50\%, thus justifying a 2:1 tapering for these types of workloads. For the large 32x32 Hx4Mesh all the boards on the same column (or on the same row) can be connected through a single 64-ports switch, thus not requiring a fat tree. Moreover, when the algorithm tries to allocate jobs to improve the communication locality, the traffic on the upper levels drops down to less than 25\% for Hx4Mesh, meaning that high tapering would not reduce performance. We see in Figure~\ref{fig:alloc-improvements} that the locality allocation heuristic does not reduce overall system utilization.
	
\begin{figure}[h!]
		\begin{center}
			\includegraphics[width=\columnwidth]{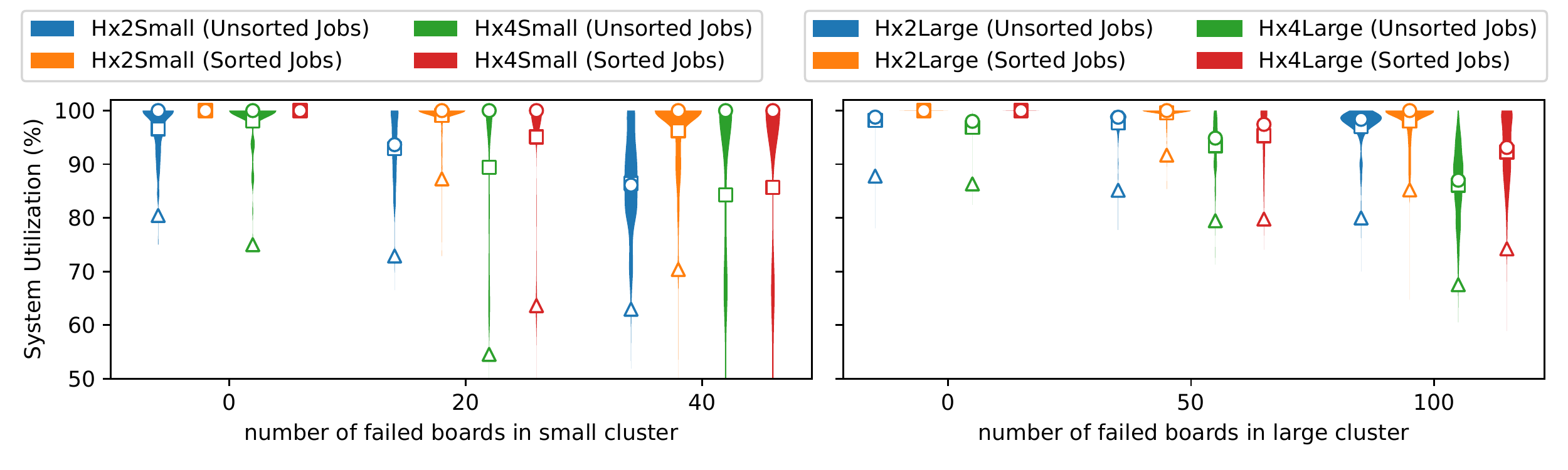}
			
		\end{center}
		\caption{HxMesh utilization for different numbers of failures.}
		\label{fig:alloc-faults}
		\vspace{-0.5em}
	\end{figure}

	Last, we investigate how random failures reduce the system utilization (due to fragmentation). Figure~\ref{fig:alloc-faults} repeats our allocation experiment but now with a varying number of randomly failed boards. We report the system utilization as the number of non-faulted boards allocated to jobs.
	
	The left shows the small clusters (256-board Hx2Mesh and 64-board Hx4Mesh) and the right shows the large clusters (4,096-board Hx2Mesh and 1,024-board Hx4Mesh). 
	In almost all the cases, our allocation algorithm achieves a median utilization of working boards (white dot inside the violin) higher than 70\%. In particular, this also holds with 40 failed boards, 62\% of the small Hx4Mesh. We also observe that meshes with fewer boards are more affected by random failures. Last, we can see that allocating the jobs in their random arrival order (rather than ordered by their size) decreases the utilization at most by 10\% on large networks.

	\subsection{Routing on HammingMesh}\label{sec:routing}
	On HxMesh, packets are routed adaptively along all shortest paths. We assume packet-level adaptive routing as implemented in High-Performance Interconnects such as Slingshot~\cite{sensi-slingshot} or InfiniBand~\cite{mlxadaptive}. To minimize packet reordering, flowlet-level adaptive routing can be used as an approximation on Ethernet networks~\cite{sinha2004harnessing}. Without loss of generality, we assume input buffered switches and credit-based flow control. We describe first how packets are routed when both source and destination accelerators are on the same board, and then how to route packets when source and destination are on two different boards.
	
	\subsubsection{Routing on the same board}
	If both source and destination are on the same board, we use adaptive routing on the torus network: The algorithm chooses the least loaded output port at each accelerator (4x4 switch) along all shortest paths between the two endpoints. Packets may be routed through fat tree switches (similar to what happens in a 2D torus) or only through on-board links.

	\subsubsection{Routing between different boards}
	If the source and destination boards are on the same row, they are adaptively routed in the source board to the closest edge (\textit{west} or \textit{east}), and then on the fat tree to the destination board's port closest to the destination. 
	On the fat tree, packets are forwarded using up/down adaptive routing~\cite{rodeheffer1991autonet}.  
	Once in the destination board, the packet is adaptively routed to the destination. It is worth noting that the packet might also need to be forwarded \textit{south} or \textit{north} within the source or destination board to reach the destination. Whether this happens in the source or destination board (or in both) depends on the local load.
	When the source and destination boards are on the same HxMesh column a similar process is applied. 
	
	If source and destination boards are on different rows and columns, packets must be forwarded through an intermediate board. 
	This board needs to be in the same row of the source board and in the same column of the destination board (or vice-versa). 
	The path selection is adaptive and minimal and packets cross two fat trees, one in each dimension.

	\subsubsection{Deadlock freedom}
	To guarantee deadlock-freedom within the board, packets are forwarded using north-last routing~\cite{753324}. Thus, the \textit{north} direction can only be taken by switches on the same column of the destination board. When forwarding packets in the tree, up/down routing guarantees deadlock-freedom. Although both the mesh and the fat tree are deadlock-free, combining the two may result in deadlocks. 
	
	A simple solution to guarantee deadlock freedom uses multiple virtual channels by increasing the virtual channel at each hop. However, this requires a number of virtual channels equal to the network diameter. Instead we observe that, because both the board and the fat tree guarantee deadlock freedom, it is enough to increase the virtual channel when jumping from one board to another (e.g., when a board injects a packet in the fat tree). Because each packet crosses at most two fat trees, this requires at most three virtual channels.

	\section{Results and Comparison}\pages{2}\label{sec:res}
	
	We now evaluate HxMesh topology options in comparison with all topologies listed in Table~\ref{tab:overview}.
	We use the Structural Simulation Toolkit (SST~\cite{SST}), a packet-level network simulator, which has been validated against the Cray Slingshot interconnect~\cite{sensi-slingshot}. SST enables us to run (slightly modified) full MPI applications directly in the simulation environment where \emph{they react to dynamic network changes (e.g., congestion)}. In total, we ran simulations of more than 120 billion packets using more than 0.6 million core hours with parallel simulations. 
	The detailed configuration is described in Appendix~\ref{app:sst}.
	We select various representative microbenchmarks and scenarios for deep learning jobs and \emph{publish the full simulation infrastructure such that readers can simulate their own job setup}.

	\subsection{Microbenchmarks}\label{sec:microbench}
	
	We start by analyzing well-known microbenchmark traffic patterns to assess and compare achievable peak bandwidth. 
	
	\subsubsection{Global traffic patterns}
	
	We first investigate global traffic patterns such as alltoall and random permutations as global-traffic workloads. We note that HammingMesh is not optimized for those patterns as they are rare on deep learning traffic.
	
	\paragraph{Alltoall}
	
	Alltoall sends messages from each process to all other processes. In our implementation, each of the $p$ processes performs $p-1$ iterations. In each iteration $i$, process $j$ sends to process $j+i \mod p$ in a balanced shift pattern.
	
\begin{figure}[h!]
			\vspace{-.7em}
			\begin{center}
				\includegraphics[width=\columnwidth]{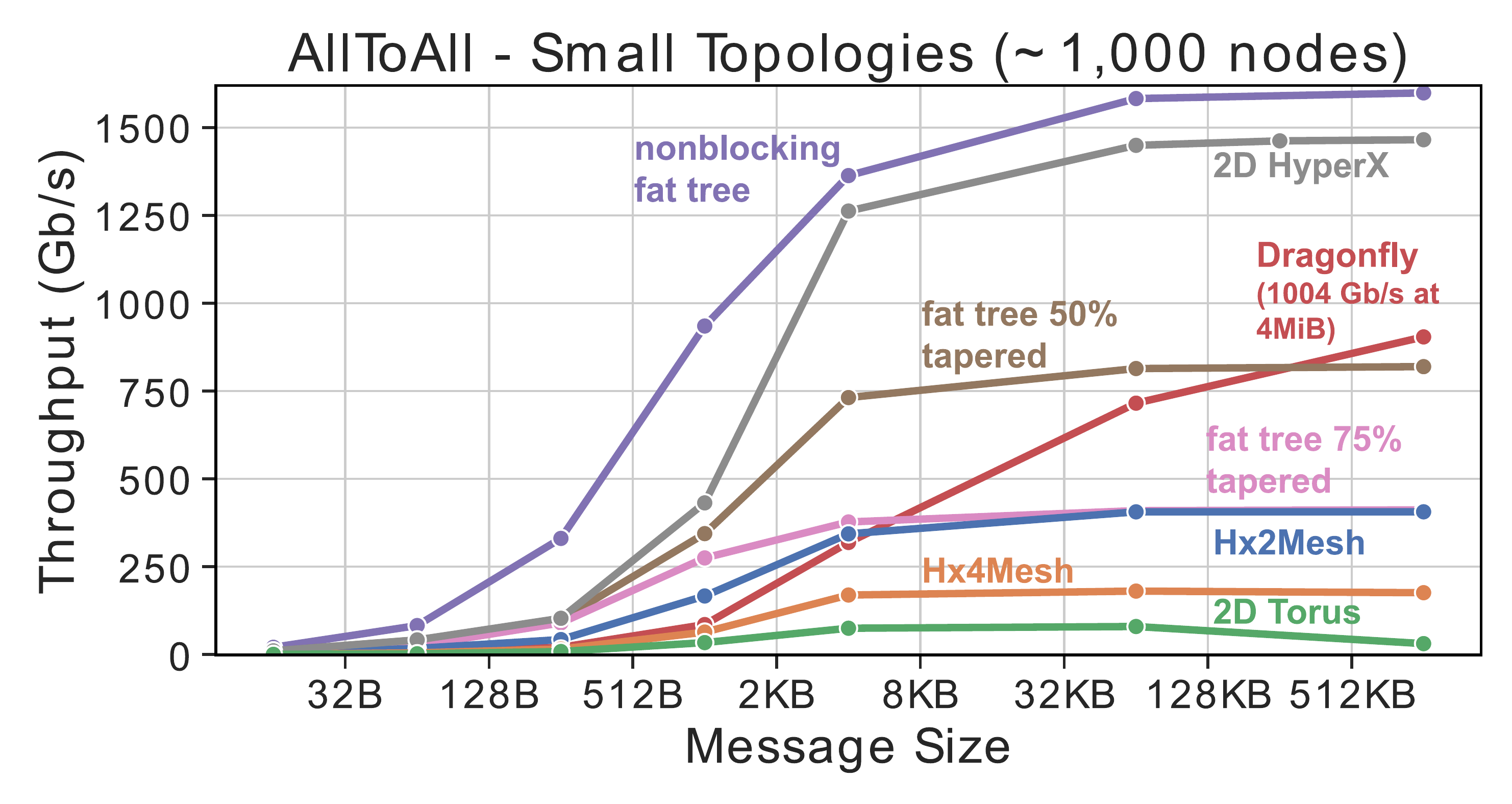}
			\end{center}
			\vspace{-1.0em}
			\caption{Alltoall on the small topologies.}
			\vspace{-1em}
			\label{fig:alltoallsmall}
	\end{figure}

	Table~\ref{tab:overview} shows the results for $1$ MiB messages while Figure~\ref{fig:alltoallsmall} shows the global bandwidth at different message sizes. Small Hx2 and Hx4Meshes achieve bandwidths around the cut width of 1/4 and 1/8, respectively (cf.~Section~\ref{sec:bisection}). This is because not all global traffic crosses the bisection cuts, especially for smaller clusters. The large cluster configuration performs closer to those bounds and loses some bandwidth due to adaptive routing overheads. Despite its lower bandwidth, even large HxMeshes remain competitive in terms of cost-per global bandwidth and some are even more cost effective on global bandwidth than fat trees.
	
	\paragraph{Random permutation}
	
\begin{figure}[h!]
			\vspace{-.9em}
			\begin{center}
				\includegraphics[width=\columnwidth]{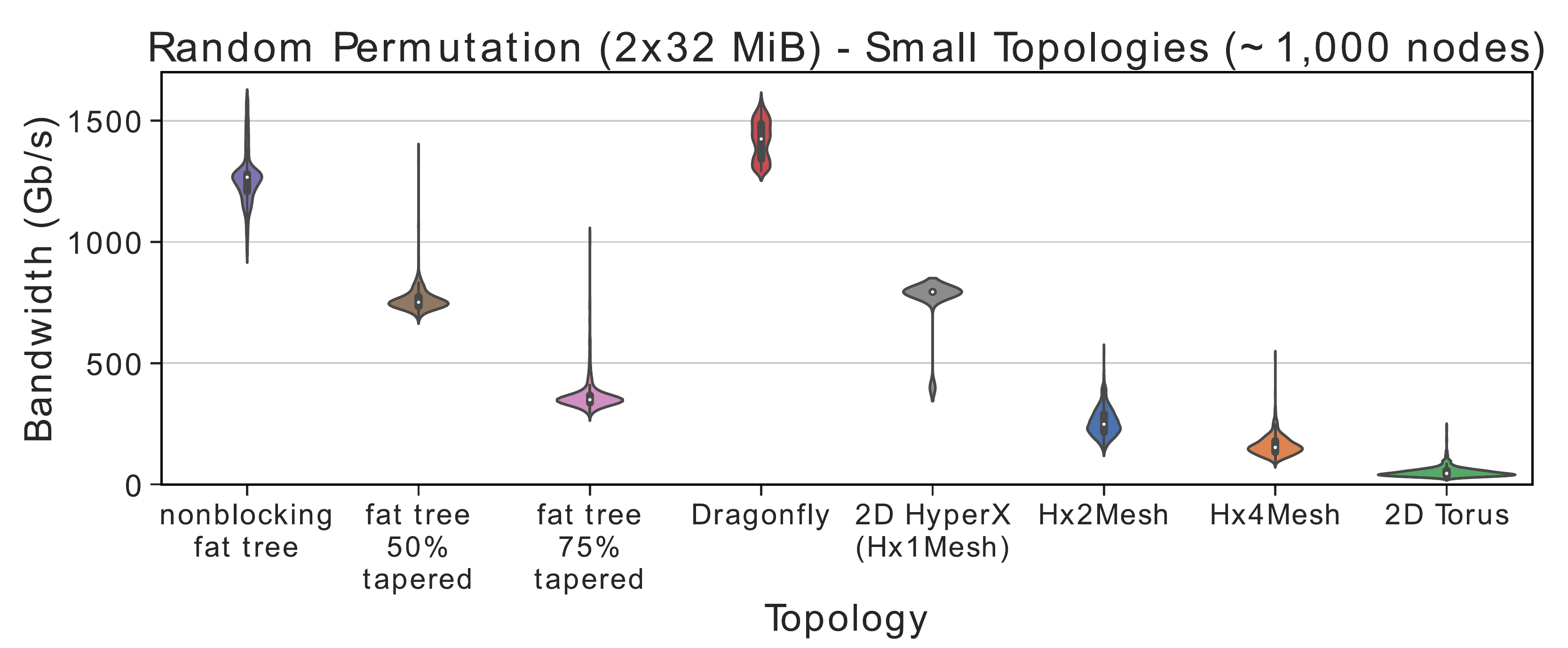}\vspace{-.8em}
			\end{center}
			\vspace{0.25em}
			\caption{Bandwidth distribution per accelerator.}
			\vspace{-.1em}
			\label{fig:randperm}
	\end{figure}

	In permutation traffic, each accelerator selects a unique random peer to send to and receive from. Here, the achieved bandwidth also depends on the location of both peers. Figure~\ref{fig:randperm} shows the distributions of receive bandwidths across all of the 1k accelerators in the small cluster configurations. 

	The figure also shows the average bandwidth as well as the the cost per average bandwidth relative to a nonblocking fat tree in the top part. 
	Our results indicate that all topologies have significant variance across different connections, which makes job placement and locality significant. HxMeshes are among the most cost effective topologies.

	\subsubsection{Reduction traffic patterns}\label{sec:fastred}
	
	We first describe our implementation of allreduce, enabled by the associativity of addition. 
	We distinguish three fundamental algorithm types: trees, pipelines, and near-optimal full-global bandwidth algorithms. 
	
	\paragraph{Simple trees}
	For small data, simple binary or binomial tree reductions are the best choice. They perform a reduction of $S$ bytes on $p$ processors in time $T\approx \log_2(p)\alpha +\log_2 (p)S\beta$\footnote{with $\approx$, we omit additive constants and minor lower-order terms for clarity}.
	This algorithm sends each data item a logarithmic number of times. It is thus inefficient for the large data sizes in deep learning training workloads and we do not consider trees in this work.
	
	\paragraph{Pipelined rings}
	With a single network interface, large data volumes can be reduced in a simple pipelined ring. Here, the data at each process is split into $p$ segments. 
	The operation proceeds in two epochs and $p-1$ rounds per epoch. In the first reduction epoch, each process $i$ sends segment $i$ to process $i+1 \mod p$ and receives a segment from process $i-1 \mod p$. The received segment is added to the local data and sent on to process $i+1 \mod p$ in the next round. After $p-1$ such rounds, each process has the full sum of one segment. The second epoch is simply sending the summed segments along the pipeline. 
	The overall time $T_p\approx 2p\alpha + 2S\beta$ is bandwidth optimal because each process only sends and receives each segment twice~\cite{optreduce}. 
	
	We propose \emph{bidirectional pipelined rings} to utilize two network interfaces by splitting the data size in half and sending each half along a different direction. The latency stays unchanged because each segment travels twice through the whole ring but the data is half in each direction, leading to a runtime of $T_{bp}\approx 2p\alpha + S\beta$. Here and in the following, $\beta$ is the time per Byte of each interface, i.e., a system with $k$ network interfaces can inject $k/\beta$ Bytes per second.
	
	We now extend this idea to four network interfaces per HxMesh plane: we use two bidirectional rings, each reducing a quarter of the data across \emph{all} accelerators. 
	The two rings are mapped to two disjoint Hamiltonian cycles covering all accelerators of the HxMesh~\cite{DBLP:journals/ijmmsc/BaeAB04}. The overall time for this scheme is $T_{rings}\approx 2p\alpha + \frac{S}{2}\beta$.
	
	\paragraph{Two-dimensional torus}
	
	Pipelined rings are bandwidth-optimal if they can be mapped to Hamiltonian cycles on the topology. However, we find that for large HxMeshes and moderate message sizes, the latency component can become a bottleneck. 
	We thus define another algorithm that uses a 2D toroidal communication pattern with $\sqrt{p}$ latency and good bandwidth usage:
	each process executes first a reduce-scatter with the other processes on the same row (cost $\sqrt{p}\alpha + \frac{S}{2}\beta$)~\cite{8868152}. Then each process runs an allreduce with the other processes on the same column, on the previously reduced chunk of size $\frac{S}{\sqrt{p}}$ (cost $2(\sqrt{p}\alpha + \frac{S}{2\sqrt{p}}\beta)$) and, eventually, an allgather with the other processes on the same row (cost \mbox{$\sqrt{p}\alpha + \frac{S}{2}\beta$}). To use all four network interfaces at the same time, two of these allreduce can be executed in parallel, each on half of the data each (one reduce consider a transposed network). Thus, the overall time for this scheme is $T\approx 2\cdot2\sqrt{p}\alpha + S\beta(\frac{1+2\sqrt{p}}{4\sqrt{p}})$.

	\paragraph{Summary} The pipeline ring and 2D torus algorithms have sparse communication patterns: each process only communicates with two or four direct neighbors that can be mapped perfectly to HxMesh.
	Broadcast and other collectives can be implemented similarly (e.g., as the second part of our allreduce) and follow similar tradeoffs.
	Furthermore, each dimension of a logical job topology is typically small as the total number of accelerators is the product of all dimensions. 
	For example, even for a very large system with 32,768 accelerators, each of the dimensions could only be of size 32 if we decompose the problem along all dimensions. 
	This means that the largest allreduce or broadcast would only be on 32 processes where ring algorithms would perform efficiently.

	\paragraph{Full system allreduce job}
	
\begin{figure}[h!]
			\begin{center}
				\vspace{-1.3em}
				\includegraphics[width=\columnwidth]{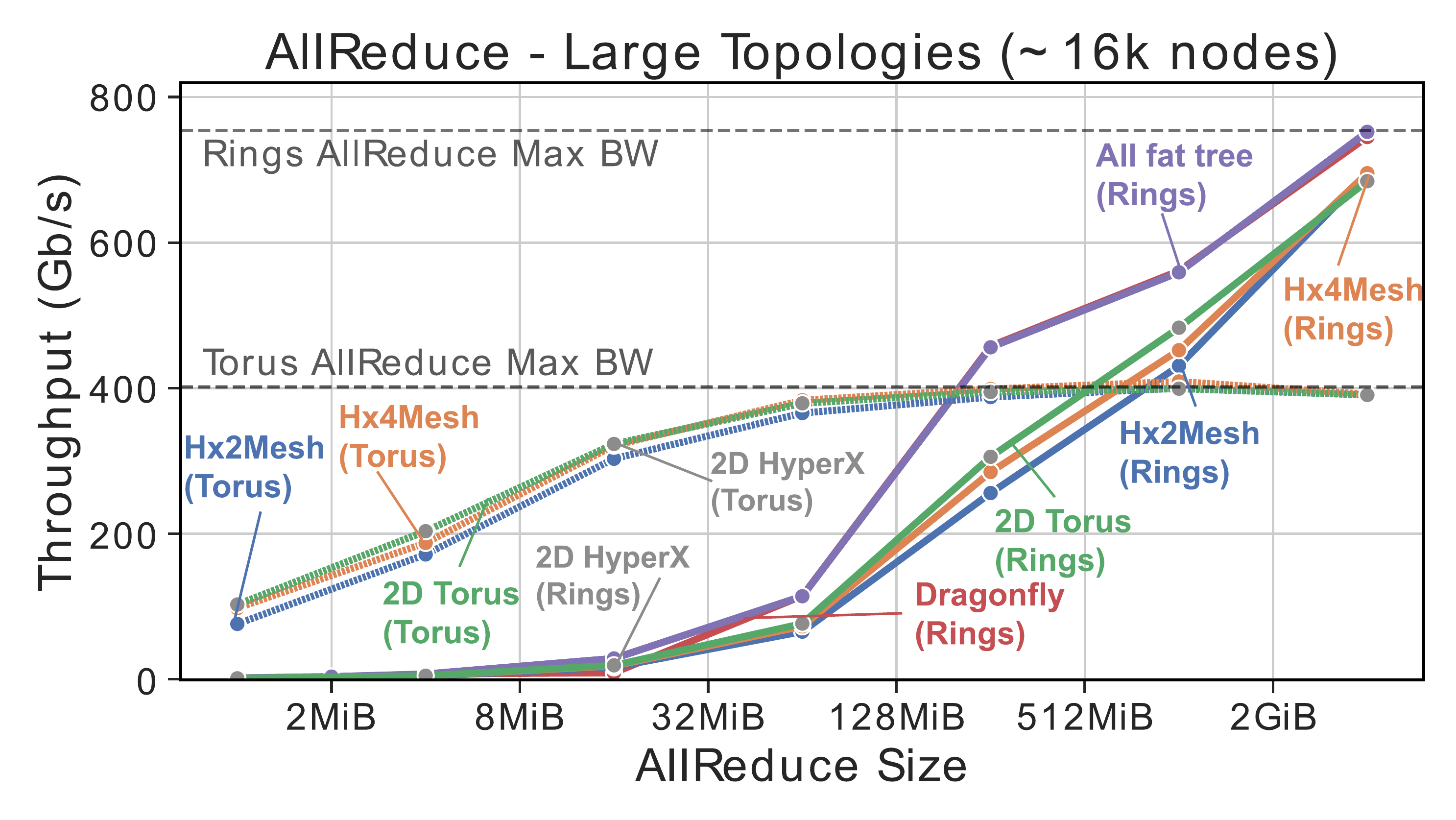}
			\end{center}
			\vspace{-1.0em}
			\caption{Global allreduce using different algorithms.}
			\vspace{-.1em}
			\label{fig:allredlarge}
	\end{figure}

	This experiment shows a single job using the last two allreduce algorithms on various topologies. 
	In Dragonfly and fat tree, each accelerator connects with a single NIC to each of the four planes and we use the standard ``ring'' algorithm.
	For the single allreduce on the large HxMesh clusters, we use both the two bidirectional rings (``rings'') as well as the two-dimensional torus (``torus'') algorithm.
	Figure~\ref{fig:allredlarge} shows the achieved bandwidths.
	
	We see that all topologies deliver nearly full bandwidth for the ring algorithms. 
	For large messages, HxMesh is 2.8x to 14.5x cheaper per bandwidth than a nonblocking fat tree (Table~\ref{tab:overview}).
	The torus algorithm, which is 2x less bandwidth-efficient, achieves higher throughput at smaller message sizes. This illustrates that multi-algorithms should be used to optimize performance, similar to established practice in MPI~\cite{thakur}.

	\subsection{DNN Workloads}\pages{2}
	We now proceed to define accurate communication patterns including computation times for real DNN models. For this, we choose three large representative models: ResNet-152, CosmoFlow, and Transformers (GPT-3) trained in FP32.
	We use NVIDIA's A100 GPU to benchmark runtimes of operators and we model communication times based on the data volumes.

	\subsubsection{Communication traffic characterization}
	
	All example models are constructed of a sequence of identical layers containing multiple operators. Each parallel dimension carries a different volume, depending on the details of the model, training hyperparameters, and the other dimensions. 
	We assume the most general case where the network can utilize all three forms of parallelism running on $D\times P\times O$ accelerators.

	\paragraph{Data dimension} If we only have data parallelism ($O=P=1$), then each process needs to reduce all gradients. 
	If we distribute the model between $O$ or $P$ dimension processes, then the 
	total allreduce size is $V_D=\frac{WN_\mathcal{P}}{OP}$. 
	The reduction happens once at the end of each iteration after processing a full minibatch and draining the pipeline. It can be overlapped per layer using nonblocking allreduce~\cite{nbc,isca_overlap,nik_alu}. 
	
\begin{figure}[h!]
			\begin{center}
				\vspace{-.7em}
				\includegraphics[width=\columnwidth]{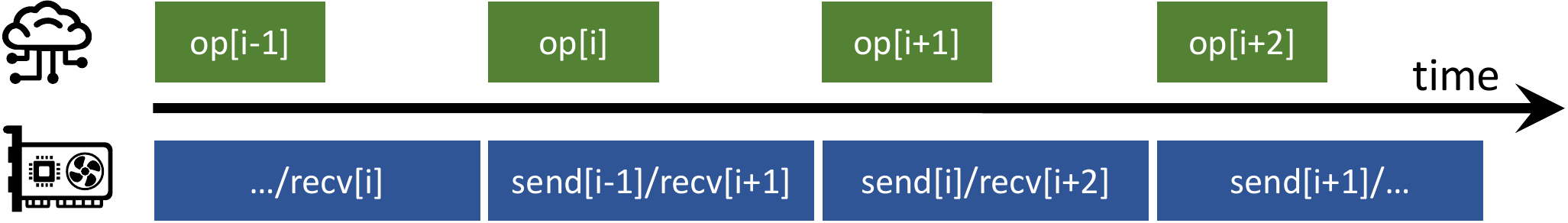}
			\end{center}
			\caption{Overlap in pipelined-parallel execution}
			\vspace{-.7em}
			\label{fig:pipesilly}
			\vspace{-.0em}
	\end{figure}

	\paragraph{Pipeline dimension} If we only have pipeline parallelism $(D=O=1)$ and $N_\mathcal{A}$ output activations at the ``cut'' layer then each process sends all $\frac{M}{P}N_\mathcal{A}$ output values to the next process in the forward pass and the same volume of errors during the backward pass. If the layer and its inputs and outputs are distributed to $O$ PEs, then the total send volume in this dimension is $V_P=\frac{MWN_\mathcal{A}}{DPO}$. 
	This communication can be hidden at each accelerator as shown in Figure~\ref{fig:pipesilly} by overlapping nonblocking send/receive operations (bottom, blue) with operator computation (top, green). 
	
	\paragraph{Operator dimension} For operator parallelism, each process' send volume depends only on the operator parallelization itself and is not influenced by either $D$ or $P$. The operator can be seen as the ``innermost loop'' in this sense. 
	Each operator distribution scheme will have its own characteristics that we capture by $V_O=WN_\mathcal{O}$.
	The operator communication volume during each forward and backward pass is a function of the local minibatch size $M/DP$ per process.

	\subsubsection{ResNets}
	ResNets achieve state of the art in many vision tasks. The standard ResNet on the ImageNet dataset has relatively small operators acting on 224$\times$224$\times$3 input data that does not warrant parallelization. Thus, we use only data parallelism ($P$=1, $O$=1). We simulate three sizes $D=\{256,512,1024\}$ with most-square jobs (e.g., 16$\times$16) with a minibatch size of $M=32,768$~\cite{you2017large} and $M/D$ examples at each accelerator.
	To reduce the latency overhead and overlap communication with computation, we divide the gradients for the 60.2M parameters of our ResNet-152 into 10 equal-sized groups. Once a group is ready, we use a nonblocking \textit{allreduce} to perform the reduction asynchronously.
	
	The compute time in one training iteration of ResNet-152 is 108 ms on 1,024 A100s. The communication can be overlapped nearly completely on all topologies. On HxMeshes and torus, the complete iteration (including overlapped communication) finishes in 110.1 ms and on the other topologies in 109.7 ms, i.e., less than 2.5\% communication overhead in the worst case. Other sizes had even less communication overhead.
	
	Thus, the effective network cost savings of HxMesh compared to other topology options can be calculated easily from Table~\ref{tab:overview}. For example, an Hx4Mesh is more than 4.1x less expensive than an 75\% tapered fat tree and more than 7.8x less expensive than a nonblocking fat tree.

	\subsubsection{CosmoFlow}

	CosmoFlow~\cite{mathuriya2018cosmoflow} is a convolutional network with large input data, one input sample of CosmoFlow is of size 128$\times$128$\times$128$\times$4. Thus, we model a hybrid of operator and data parallelism for CosmoFlow. We execute with $D=256$, $P=1$, $O=4$ on 1,024 accelerators. We assume a minibatch size of $M=8,192$~\cite{mathuriya2018cosmoflow} with a local batch size of 32. CosmoFlow mainly consists of convolutional layers and fully-connected layers, with $8.9M$ trainable parameters total. In the forward pass, each convolutional layer has to exchange a halo region of the input data with its neighbors using send/recv communication; each fully-connected layer uses allgather to collect the input data. The backwards pass is similar using send/recv, reducescatter, and allreduce communications.
	
	CosmoFlow is more complex with two levels of parallelism and different communication operations. The compute time of CosmoFlow is 44.3 ms on A100. Nearly all communication time can be overlapped leading to less than 2\% overhead on all topologies but Hx4Mesh and torus (3.4\% and 4.4\%).

	\subsubsection{DLRM}
	
\begin{figure*}[ht!]
		\vspace{-1.7em}
		\begin{center}
			\includegraphics[width=0.99\textwidth]{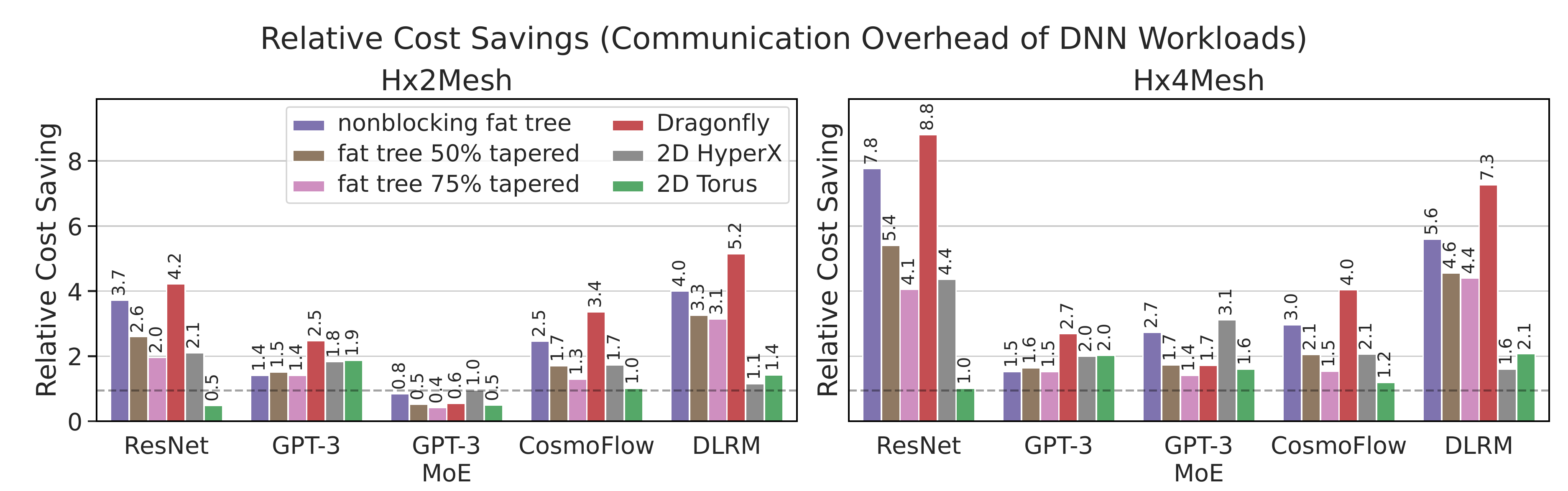}\vspace{-.8em}
		\end{center}
		\vspace{.2em}
		\caption{HxMesh cost savings relative to other topologies.}
		\vspace{-1.3em}
		\label{fig:dnnsavings}
		\vspace{-0.3em}
\end{figure*}

	DLRM~\cite{yang2020training, naumov2019deep, gupta2020architectural} uses a combination of model parallelism and data parallelism for its embedding and MLP layers, respectively. Two alltoall operations aggregate sparse embedding lookups in the forward pass, and their corresponding gradients in backward pass. Allreduce is required to synchronize the gradients of the data-parallel MLP layers. The parallelism of DLRM is limited by both the mini-batch size and the embedding dimension. DLRM is trained with up to 128 GPU nodes~\cite{yang2020training, mudigere2021software}. The total runtimes on the  fat tree variants are 2.96 ms, 2.97 ms, and 2.99 ms, respectively. On torus, the code executes for 3.12 ms. HyperX is at 2.94 ms. Hx2Mesh and Hx4Mesh are at 2.97 ms and 3.00 ms, respectively. 
		On A100, DLRM computes around 95 us, 209 us, and 796 us for the embedding, feature interaction, and MLP layers respectively, and communicates 1 MB per alltoall and 2.96 MB per allreduce.
		All simulation results are shown in Figure~\ref{fig:dnnsavings}.

	\subsubsection{Transformers}
	Transformers are the most communication intensive~\cite{data-movement-is-all-you-need}. A transformer block consists of multi-head attention (MHA) and two feed-forward (FF) layers. The MHA and FF input/outputs are of size (embedding dimension$\times$batch$\times$sequence length).
	For example, GPT-3's~\cite{brown2020language} feed forward layers multiply 49,152$\times$12,288 with 12,288$\times$2,048 matrices per example in each layer.
	
	GPT-3 has a total of 96 layers and each layer has activations of size $N_\mathcal{A}=4\cdot 2,048\times 12,288\approx 100$MB
	per example as input and output. We choose $P=96$, such that each pipeline stage processes one layer, and no data parallelism ($D=1$). 
	For operator parallelism, we use $O=4$ and the scheme outlined by Megatron-LM~\cite{shoeybi2020megatronlm}, which  performs one allreduce for FF and one for MHA in both the forward and backward passes. 
	
	All operations are the same size as the layer input/output.
	Thus, the volume for both pipeline communication and operator-dimension allreduce is $N_\mathcal{A}$ per example for forward and backward passes.
	One iteration of GPT-3 computes for 31.8 ms. The total runtimes on the three fat tree variants are 34.8 ms, 36.4 ms, and 37.5 ms, respectively. On torus, the code executes for 72.2 ms per iteration. HyperX is at 40.9 ms. Hx2 and Hx4Mesh are at 41.7 ms and 49.9 ms, respectively.

	For GPT-3 with Mixture-of-Experts (MoEs)~\cite{gshard}, we use 16 experts. In GPT-3, the FFs have $1.8B$ parameters. Therefore, each expert has $1.8B / 16 \approx 113M$ parameters. MoEs perform two alltoalls for FF in both the forward and backward passes, and all operations are the same size as the input/output. The computation time on an A100 is 49.9 ms. The total runtime on the fat trees varies from 52.2 ms to 52.9 ms depending on tapering. On torus, the code executes for 73.8 ms per iteration. HyperX takes 53.9 ms while Hx2 and Hx4Mesh are at 58.3 ms and 63.3 ms, respectively. 
	
	Figure~\ref{fig:dnnsavings} shows the relative cost savings of HxMesh compared to other topologies. These are calculated as the ratio of the network costs  in Section~\ref{tab:overview} times the inverse of the ratio of communication overheads presented in this section.

	We conclude that both Hx2 and Hx4Mesh significantly reduce network costs for DNN workloads. While some torus network configurations can be cheaper than Hx2Mesh, they provide significantly less allocation and management flexibility, especially in the presence of failures. Moreover, we also conclude that even in the presence of alltoall communications patterns in GPT-3 MoE and DLRM HxMesh topologies still offer a significant cost advantage compared to traditional topologies. As the scale of the network increases, Hx4Mesh becomes significantly more cost efficient than Hx2Mesh especially in the presence of alltoall traffic.

	\vspace{-0.5em}
	\section{Related Work}
	
	Dragonfly~\cite{DF} and PERCS~\cite{ibm-percs-network} are global bandwidth topologies designed for lowest cost given electrical and optical cables and highest global bandwidth. Slim Fly~\cite{SF} and other diameter-2 topologies~\cite{diameter-2-topos} provide the maximum number of nodes at diameter two (number of switch hops) and full global bandwidth. While they minimize the number of switches and cables, their structure does not take advantage of technology parameters such as PCB traces or DAC vs. AoC cables. 
	
	Other closely related two-dimensional topologies are Flattened Butterfly~\cite{flatfly} and 2D Hyper-X networks~\cite{hyperx}. Kim et al.~\cite{DF} show that Dragonfly topologies are 20\% more cost effective than Flattened Butterflies and HxMesh improves over both. 2D Hyper-X topologies are isomorphic to 2D Hamming graphs, i.e., Hx1Meshes with a 1x1 board layout. The Cube Collective topology~\cite{keithcc} uses on-board links in combination with a fully-connected global topology similar to a Dragonfly.
	
	Large-scale 3D torus networks have been used in supercomputers~\cite{BW,BGL}. Google's TPUs also utilize torui for cost-efficiency~\cite{tpusc}. Higher-dimensional torus networks provide higher global bandwidth~\cite{fugakunet}. Yet, torus networks remain relatively inflexible for allocating jobs and dealing with failures.
	
	A specific co-design between the network architecture and the allreduce algorithm has been proposed~\cite{9499662} but it lacks the cost-effectiveness of HammingMesh, especially on a more general scenario involving also alltoall operations. Moreover it is relatively more complex and it also lacks a robust job failures analysis. 
	
	BML~\cite{BML} uses the BCube topology to take advantage of multiple NICs per server and implement only the data-parallel dimension (allreduce). It achieves about twice the performance for a ring algorithm compared to a parameter server on a nonblocking fat tree and it can be built with 40\% of the switches of the fat tree. We show that an Hx4Mesh is more cost efficient by achieving the optimal ring bandwidth at 10.8\% (1/9.3x) of the cost. We also show how model (pipeline and operator) parallelism can be mapped efficiently for complex parallel models.
	
	HammingMesh explicitly combines on-board inexpensive PCB-based mesh topologies with global AoC and DAC cabled topologies. This general idea opens many avenues for future designs such as combining different board and global topologies to accommodate technology trends such as larger packages with chiplets or co-packaged optics.

	\section{Discussion and extensions}
	
	The design of HammingMesh is based on two fundamental insights: (1) deep learning workloads do not require full global bandwidth and exhibit torus-like communication patterns and (2) combining local low-cost meshes with a global full-bandwidth topology unlocks a design space with intriguing tradeoffs between cost, flexibility, and local/global bandwidth. 
	We show examples from deep learning but we expect that other similar workloads such as tensor contractions, relevant in quantum system simulation~\cite{haner-sim} or (multi)linear algebra~\cite{greg-lu} can also be supported efficiently.
	
	We could only show a small part of the design space, combining local 2D meshes with fat trees. Straight-forward extensions are a mix of different local torus topologies and other global layouts (e.g., 1D HxMeshes). Furthermore, one could consider different global-bandwidth topologies, such as Dragonfly or Slim Fly to connect the rows and columns. Such topologies enable connecting more nodes cheaply and could even span multiple (potentially all) rows and/or columns. 
	
	Much of  HxMesh's cost benefit comes from cheaper global topologies. One could change the arrangement of endpoints in racks to enable more aggressive use of cheaper copper cables. 
	
	Several works shown the performance advantages of offloading the allreduce operation to switches. The idea behind these works is to build a tree where the leaves are the compute nodes, and the intermediate nodes are a subset of switches in the network. Data packets get reduced by the switch and, once they reach the root of the tree, the reduced data is multicasted back to the nodes down the tree. This is orthogonal to the topology design and similar solutions can be applied to HammingMesh topologies, since it is enough to build such a reduction tree. Switches in the fat tree can use SHARP~\cite{sharp,sharp2} rely on programmable switches~\cite{flare,switchml}, whereas a similar technology can also be adopted within the boards, as recently shown for systems similar to DGX-2~\cite{nvidiaallreduce}. 
	
	Existing state distribution strategies such as ZeRO~\cite{zero} or parallelizing frameworks such as FlexFlow~\cite{flexflow} can take advantage of HammingMesh topologies and achieve highest performance when the traffic is mapped to rings or the collective primitives discussed above.

	\section{Conclusions}
	
	HammingMesh is optimized specifically for machine learning workloads and their communication patterns. It relies on the observation that deep learning training uses three-dimensional communication patterns and rarely needs global bandwidth. It supports extreme local bandwidth while controlling the cost of global bandwidth.
	It banks on an inexpensive local PCB-mesh interconnect together with a workload-optimized global connectivity forming virtual torus networks at adjustable global bandwidth.  
	
	Due to the lower number of switches and external cables, it can be nearly always more cost effective than torus networks while also offering higher global bandwidth and significantly higher flexibility in job allocation and dealing with failures.
	
	All-in-all, we believe that HammingMesh will drive future deep learning systems and will also support adjacent workloads, such as (multi)linear algebra, quantum simulation, or parallel solvers, that have Cartesian communication patterns.
	
	\section*{Acknowledgment}
		We thank Microsoft for hosting TH's sabbatical where much of the idea was developed~\cite{hxnet,cglass}. 
		We thank the whole Azure Hardware Architecture team and especially Doug Burger for their continued support and deep technical discussions.
		We thank the Swiss National Supercomputing Center (CSCS) for the compute resources on Piz Daint and the Slim Fly cluster (thanks to Hussein Harake) to run the simulations.
		\begin{wrapfigure}{r}{1.4cm}
		    \vspace{-0.4cm}
		    \includegraphics[width=1.4cm]{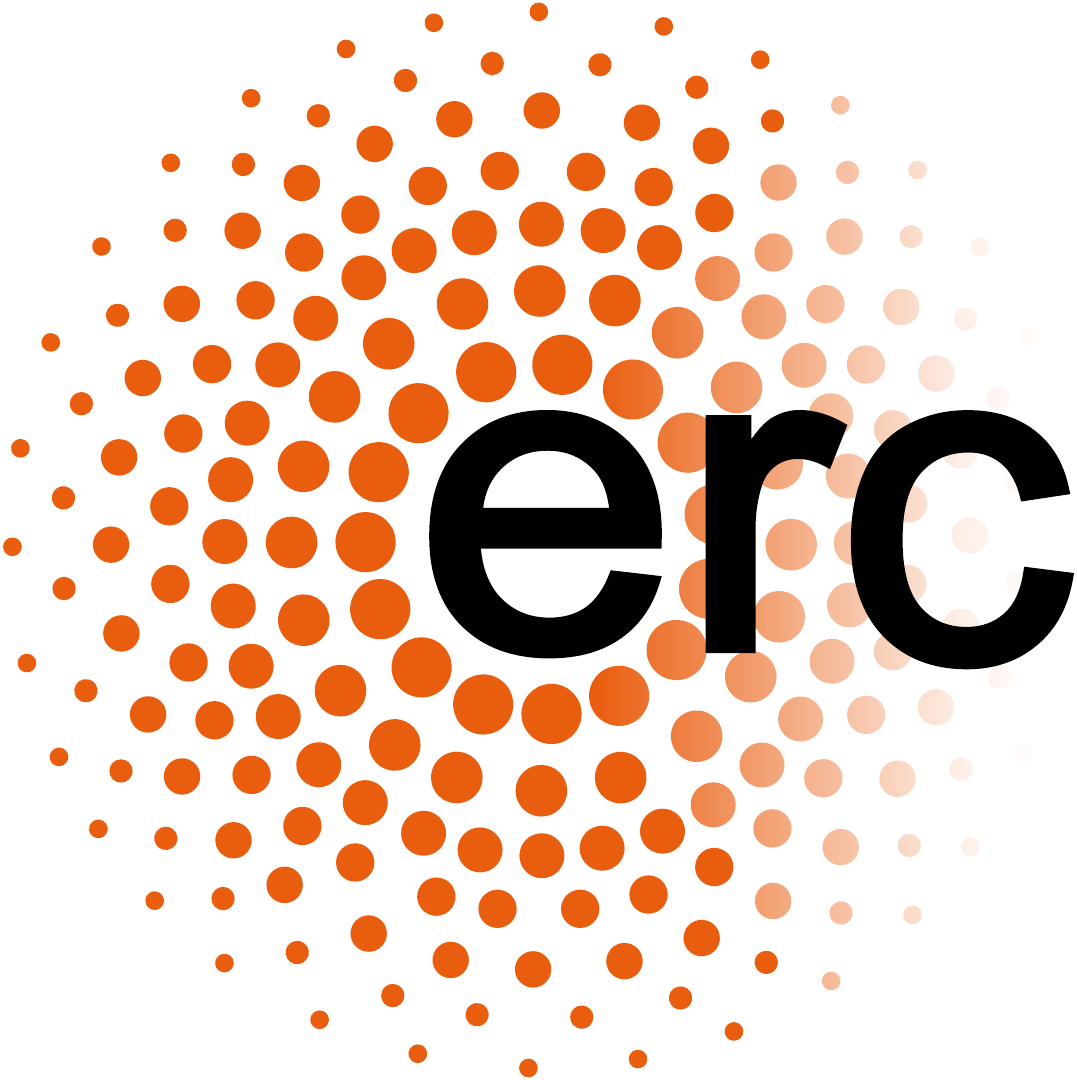}
		\end{wrapfigure}
		This project has received funding from the European Research Council under grant agreement PSAP, No. 101002047. Daniele De Sensi is supported by an ETH Postdoctoral Fellowship (19-2 FEL-50).

	\bibliographystyle{ieeetran}
	\bibliography{hammingnet}

\begin{thebibliography}{10}
\providecommand{\url}[1]{#1}
\csname url@samestyle\endcsname
\providecommand{\newblock}{\relax}
\providecommand{\bibinfo}[2]{#2}
\providecommand{\BIBentrySTDinterwordspacing}{\spaceskip=0pt\relax}
\providecommand{\BIBentryALTinterwordstretchfactor}{4}
\providecommand{\BIBentryALTinterwordspacing}{\spaceskip=\fontdimen2\font plus
\BIBentryALTinterwordstretchfactor\fontdimen3\font minus
  \fontdimen4\font\relax}
\providecommand{\BIBforeignlanguage}[2]{{%
\expandafter\ifx\csname l@#1\endcsname\relax
\typeout{** WARNING: IEEEtran.bst: No hyphenation pattern has been}%
\typeout{** loaded for the language `#1'. Using the pattern for}%
\typeout{** the default language instead.}%
\else
\language=\csname l@#1\endcsname
\fi
#2}}
\providecommand{\BIBdecl}{\relax}
\BIBdecl

\bibitem{sw20}
\BIBentryALTinterwordspacing
A.~Karpathy, ``Software 2.0,'' November 2017, [Online; posted 11-Nov-2017].
  [Online]. Available:
  \url{https://karpathy.medium.com/software-2-0-a64152b37c35}
\BIBentrySTDinterwordspacing

\bibitem{openaicompute}
D.~H. Dario~Amodei, ``Ai and compute,'' online
  https://openai.com/blog/ai-and-compute/, 05 2018.

\bibitem{kaplan2020scaling}
J.~Kaplan, S.~McCandlish, T.~Henighan, T.~B. Brown, B.~Chess, R.~Child,
  S.~Gray, A.~Radford, J.~Wu, and D.~Amodei, ``Scaling laws for neural language
  models,'' 2020.

\bibitem{data-movement-is-all-you-need}
A.~Ivanov, N.~Dryden, T.~Ben-Nun, S.~Li, and T.~Hoefler, ``{Data Movement Is
  All You Need: A Case Study on Optimizing Transformers},'' in
  \emph{Proceedings of Machine Learning and Systems 3 (MLSys 2021)}, Apr. 2021.

\bibitem{tcv100}
{NVIDIA Corporation}, ``{NVIDIA Tesla V100 GPU Architecture},'' Tech. Rep.
  WP-08608-001\_v1.1, 08 2017.

\bibitem{xilinxaie}
{Xilinx Corporation}, ``{Xilinx AI Engines and Their Applications},'' Tech.
  Rep. WP506 (v1.1) July 10, 2020, 07 2020.

\bibitem{msfp}
B.~Darvish~Rouhani, D.~Lo, R.~Zhao, M.~Liu, J.~Fowers, K.~Ovtcharov,
  A.~Vinogradsky, S.~Massengill, L.~Yang, R.~Bittner, A.~Forin, H.~Zhu, T.~Na,
  P.~Patel, S.~Che, L.~Chand~Koppaka, X.~SONG, S.~Som, K.~Das, S.~T,
  S.~Reinhardt, S.~Lanka, E.~Chung, and D.~Burger, ``Pushing the limits of
  narrow precision inferencing at cloud scale with microsoft floating point,''
  in \emph{Advances in Neural Information Processing Systems}, H.~Larochelle,
  M.~Ranzato, R.~Hadsell, M.~F. Balcan, and H.~Lin, Eds., vol.~33.\hskip 1em
  plus 0.5em minus 0.4em\relax Curran Associates, Inc., 2020, pp.
  10\,271--10\,281.

\bibitem{sze}
V.~Sze, Y.-H. Chen, T.-J. Yang, and J.~S. Emer, ``Efficient processing of deep
  neural networks: A tutorial and survey,'' \emph{Proceedings of the IEEE},
  vol. 105, no.~12, pp. 2295--2329, 2017.

\bibitem{kepner}
\BIBentryALTinterwordspacing
A.~Reuther, P.~Michaleas, M.~Jones, V.~Gadepally, S.~Samsi, and J.~Kepner,
  ``Survey of machine learning accelerators,'' in \emph{2020 {IEEE} High
  Performance Extreme Computing Conference ({HPEC})}.\hskip 1em plus 0.5em
  minus 0.4em\relax {IEEE}, sep 2020. [Online]. Available:
  \url{https://doi.org/10.1109%2Fhpec43674.2020.9286149}
\BIBentrySTDinterwordspacing

\bibitem{sparsity-in-dl}
T.~Hoefler, D.~Alistarh, T.~Ben-Nun, N.~Dryden, and A.~Peste, ``{Sparsity in
  Deep Learning: Pruning and growth for efficient inference and training in
  neural networks},'' \emph{Journal of Machine Learning Research}, vol.~22, no.
  241, pp. 1--124, Sep. 2021.

\bibitem{6634083}
P.~Kogge and J.~Shalf, ``Exascale computing trends: Adjusting to the "new
  normal"' for computer architecture,'' \emph{Computing in Science
  Engineering}, vol.~15, no.~6, pp. 16--26, 2013.

\bibitem{tpuv2}
T.~Norrie, N.~Patil, D.~H. Yoon, G.~Kurian, S.~Li, J.~Laudon, C.~Young, N.~P.
  Jouppi, and D.~A. Patterson, ``{Google's Training Chips Revealed: TPUv2 and
  TPUv3.}'' 08 2020, {Hot Chips Symposium, pp. 1-70. 2020}.

\bibitem{trainium}
P.~DeSantis, ``{Keynote at AWS re:Invent 2021},'' online
  https://www.youtube.com/watch?v=9NEQbFLtDmg\&t=4105s, 12 2021.

\bibitem{a100dgx}
{NVIDIA Corporation}, ``{NVIDIA DGX A100 System Architecture},'' Tech. Rep.
  WP-10083-001\_v01, 07 2020.

\bibitem{h100}
------, ``{NVIDIA H100 Tensor Core GPU Architecture},'' Tech. Rep. V1.01, 03
  2022.

\bibitem{dojo}
G.~Venkataramanan, ``Talk at tesla ai day,'' online
  https://www.youtube.com/watch?v=j0z4FweCy4M\&t=6775s, 08 2021.

\bibitem{switchml}
A.~Sapio, M.~Canini, C.-Y. Ho, J.~Nelson, P.~Kalnis, C.~Kim, A.~Krishnamurthy,
  M.~Moshref, D.~R.~K. Ports, and P.~Richtárik, ``{Scaling Distributed Machine
  Learning with In-Network Aggregation},'' in \emph{Proceedings of the 18th
  USENIX Symposium on Networked Systems Design and Implementation (NSDI 21)},
  Apr 2021.

\bibitem{sparcml}
C.~Renggli, D.~Alistarh, M.~Aghagolzadeh, and T.~Hoefler, ``{SparCML:
  High-Performance Sparse Communication for Machine Learning},'' in
  \emph{Proceedings of the International Conference for High Performance
  Computing, Networking, Storage and Analysis (SC19)}, Nov. 2019.

\bibitem{distdl-preprint}
T.~Ben-Nun and T.~Hoefler, ``{Demystifying Parallel and Distributed Deep
  Learning: An In-Depth Concurrency Analysis},'' \emph{ACM Comput. Surv.},
  vol.~52, no.~4, pp. 65:1--65:43, Aug. 2019.

\bibitem{top-k}
D.~Alistarh, T.~Hoefler, M.~Johansson, S.~Khirirat, N.~Konstantinov, and
  C.~Renggli, ``{The Convergence of Sparsified Gradient Methods},'' in
  \emph{Advances in Neural Information Processing Systems 31}.\hskip 1em plus
  0.5em minus 0.4em\relax Curran Associates, Inc., Dec. 2018.

\bibitem{localsgd1}
A.~Dieuleveut and K.~K. Patel, ``Communication trade-offs for local-sgd with
  large step size,'' in \emph{Advances in Neural Information Processing
  Systems}, H.~Wallach, H.~Larochelle, A.~Beygelzimer, F.~d\textquotesingle
  Alch\'{e}-Buc, E.~Fox, and R.~Garnett, Eds., vol.~32.\hskip 1em plus 0.5em
  minus 0.4em\relax Curran Associates, Inc., 2019.

\bibitem{stich2019local}
S.~U. Stich, ``Local sgd converges fast and communicates little,'' 2019.

\bibitem{gorbunov2020local}
E.~Gorbunov, F.~Hanzely, and P.~Richtárik, ``Local sgd: Unified theory and new
  efficient methods,'' 2020.

\bibitem{khaled2020tighter}
A.~Khaled, K.~Mishchenko, and P.~Richtárik, ``Tighter theory for local sgd on
  identical and heterogeneous data,'' 2020.

\bibitem{dryden-quant}
N.~Dryden, S.~A. Jacobs, T.~Moon, and B.~Van~Essen, ``Communication
  quantization for data-parallel training of deep neural networks,'' in
  \emph{Proceedings of the Workshop on Machine Learning in High Performance
  Computing Environments}, ser. MLHPC '16.\hskip 1em plus 0.5em minus
  0.4em\relax IEEE Press, 2016, p.~18.

\bibitem{NEURIPS2018_3328bdf9}
J.~Wangni, J.~Wang, J.~Liu, and T.~Zhang, ``Gradient sparsification for
  communication-efficient distributed optimization,'' in \emph{Advances in
  Neural Information Processing Systems}, S.~Bengio, H.~Wallach, H.~Larochelle,
  K.~Grauman, N.~Cesa-Bianchi, and R.~Garnett, Eds., vol.~31.\hskip 1em plus
  0.5em minus 0.4em\relax Curran Associates, Inc., 2018.

\bibitem{NEURIPS2018_31445061}
D.~Alistarh, T.~Hoefler, M.~Johansson, N.~Konstantinov, S.~Khirirat, and
  C.~Renggli, ``The convergence of sparsified gradient methods,'' in
  \emph{Advances in Neural Information Processing Systems}, S.~Bengio,
  H.~Wallach, H.~Larochelle, K.~Grauman, N.~Cesa-Bianchi, and R.~Garnett, Eds.,
  vol.~31.\hskip 1em plus 0.5em minus 0.4em\relax Curran Associates, Inc.,
  2018.

\bibitem{45187}
\BIBentryALTinterwordspacing
J.~Chen, R.~Monga, S.~Bengio, and R.~Jozefowicz, ``Revisiting distributed
  synchronous sgd,'' in \emph{International Conference on Learning
  Representations Workshop Track}, 2016. [Online]. Available:
  \url{https://arxiv.org/abs/1604.00981}
\BIBentrySTDinterwordspacing

\bibitem{huang2019gpipe}
Y.~Huang, Y.~Cheng, A.~Bapna, O.~Firat, M.~X. Chen, D.~Chen, H.~Lee, J.~Ngiam,
  Q.~V. Le, Y.~Wu, and Z.~Chen, ``Gpipe: Efficient training of giant neural
  networks using pipeline parallelism,'' 2019.

\bibitem{harlap2018pipedream}
A.~Harlap, D.~Narayanan, A.~Phanishayee, V.~Seshadri, N.~Devanur, G.~Ganger,
  and P.~Gibbons, ``Pipedream: Fast and efficient pipeline parallel dnn
  training,'' 2018.

\bibitem{chimera}
S.~Li and T.~Hoefler, ``{Chimera: Efficiently Training Large-Scale Neural
  Networks with Bidirectional Pipelines},'' in \emph{Proceedings of the
  International Conference for High Performance Computing, Networking, Storage
  and Analysis (SC21)}.\hskip 1em plus 0.5em minus 0.4em\relax ACM, Nov. 2021.

\bibitem{MLSYS2021_6c8349cc}
B.~Yang, J.~Zhang, J.~Li, C.~Re, C.~Aberger, and C.~De~Sa, ``Pipemare:
  Asynchronous pipeline parallel dnn training,'' in \emph{Proceedings of
  Machine Learning and Systems}, A.~Smola, A.~Dimakis, and I.~Stoica, Eds.,
  vol.~3, 2021, pp. 269--296.

\bibitem{prisacari-dragonfly-mapping}
B.~Prisacari, G.~Rodriguez, P.~Heidelberger, D.~Chen, C.~Minkenberg, and
  T.~Hoefler, ``{Efficient Task Placement and Routing in Dragonfly Networks
  },'' in \emph{Proceedings of the 23rd ACM International Symposium on
  High-Performance Parallel and Distributed Computing (HPDC'14)}.\hskip 1em
  plus 0.5em minus 0.4em\relax ACM, Jun. 2014.

\bibitem{SF}
M.~Besta and T.~Hoefler, ``{Slim Fly: A Cost Effective Low-Diameter Network
  Topology},'' Nov. 2014, proceedings of the International Conference on High
  Performance Computing, Networking, Storage and Analysis (SC14).

\bibitem{prisacari-ics-bandwidth-opt-alltoall}
B.~Prisacari, G.~Rodriguez, C.~Minkenberg, and T.~Hoefler, ``{Bandwidth-optimal
  All-to-all Exchanges in Fat Tree Networks},'' in \emph{Proceedings of the
  27th International ACM Conference on International Conference on
  Supercomputing}.\hskip 1em plus 0.5em minus 0.4em\relax ACM, Jun. 2013, pp.
  139--148.

\bibitem{DF}
J.~Kim, W.~J. Dally, S.~Scott, and D.~Abts, ``Technology-driven,
  highly-scalable dragonfly topology,'' in \emph{2008 International Symposium
  on Computer Architecture}, 2008, pp. 77--88.

\bibitem{diameter-2-topos}
G.~Kathareios, C.~Minkenberg, B.~Prisacari, G.~Rodriguez, and T.~Hoefler,
  ``{Cost-Effective Diameter-Two Topologies: Analysis and Evaluation}.''\hskip
  1em plus 0.5em minus 0.4em\relax ACM, Nov. 2015, in Proceedings of the
  International Conference for High Performance Computing, Networking, Storage
  and Analysis (SC15).

\bibitem{arrays-trees-hypercubes}
F.~T. Leighton, \emph{Introduction to parallel algorithms and architectures:
  Arrays, trees, hypercubes}.\hskip 1em plus 0.5em minus 0.4em\relax Elsevier,
  1991.

\bibitem{gareyjohnson}
M.~R. Garey and D.~S. Johnson, \emph{Computers and Intractability; A Guide to
  the Theory of NP-Completeness}.\hskip 1em plus 0.5em minus 0.4em\relax USA:
  W. H. Freeman \& Co., 1990.

\bibitem{276938}
\BIBentryALTinterwordspacing
``{MLaaS} in the wild: Workload analysis and scheduling in {Large-Scale}
  heterogeneous {GPU} clusters,'' in \emph{19th USENIX Symposium on Networked
  Systems Design and Implementation (NSDI 22)}.\hskip 1em plus 0.5em minus
  0.4em\relax Renton, WA: USENIX Association, Apr. 2022. [Online]. Available:
  \url{https://www.usenix.org/conference/nsdi22/presentation/weng}
\BIBentrySTDinterwordspacing

\bibitem{alibaba-trace}
\BIBentryALTinterwordspacing
Alibaba, ``Alibaba cluster trace program,'' 2020, [Online; accessed
  04-Mar-2022]. [Online]. Available:
  \url{\url{https://github.com/alibaba/clusterdata/blob/master/cluster-trace-gpu-v2020/README.md}}
\BIBentrySTDinterwordspacing

\bibitem{sensi-slingshot}
D.~De~Sensi, S.~Di~Girolamo, K.~H. McMahon, D.~Roweth, and T.~Hoefler, ``{An
  In-Depth Analysis of the Slingshot Interconnect},'' in \emph{Proceedings of
  the International Conference for High Performance Computing, Networking,
  Storage and Analysis (SC20)}, Nov. 2020.

\bibitem{mlxadaptive}
{NVIDIA Corporation}, ``{NVIDIA InfiniBand Adaptive Routing Technology},''
  Tech. Rep. WP-10326-001\_v01, 07 2020.

\bibitem{sinha2004harnessing}
S.~Sinha, S.~Kandula, and D.~Katabi, ``{Harnessing TCPs Burstiness using
  Flowlet Switching},'' in \emph{3rd ACM SIGCOMM Workshop on Hot Topics in
  Networks (HotNets)}, San Diego, CA, November 2004.

\bibitem{rodeheffer1991autonet}
T.~L. Rodeheffer, C.~Thacker, A.~Birrell, T.~Rodeheffer, H.~Murray,
  M.~Schroeder, E.~Satterthwaite, R.~Needham, M.~Burrows, M.~D. Schroeder, and
  M.~Schroeder, ``Autonet: A high-speed, self-configuring local area network
  using point-to-point links,'' \emph{IEEE Journal on Select Areas of
  Communication}, vol.~9, October 1991.

\bibitem{753324}
C.~Glass and L.~Ni, ``The turn model for adaptive routing,'' in
  \emph{Proceedings the 19th Annual International Symposium on Computer
  Architecture}, 1992, pp. 278--287.

\bibitem{SST}
\BIBentryALTinterwordspacing
H.~Adalsteinsson, S.~Cranford, D.~A. Evensky, J.~P. Kenny, J.~Mayo, A.~Pinar,
  and C.~L. Janssen, ``A simulator for large-scale parallel computer
  architectures,'' \emph{Int. J. Distrib. Syst. Technol.}, vol.~1, no.~2, p.
  5773, apr 2010. [Online]. Available:
  \url{https://doi.org/10.4018/jdst.2010040104}
\BIBentrySTDinterwordspacing

\bibitem{optreduce}
\BIBentryALTinterwordspacing
M.~Barnett, R.~Littlefield, D.~Payne, and R.~Vandegeijn, ``Global combine
  algorithms for 2-d meshes with wormhole routing,'' \emph{J. Parallel Distrib.
  Comput.}, vol.~24, no.~2, p. 191201, feb 1995. [Online]. Available:
  \url{https://doi.org/10.1006/jpdc.1995.1018}
\BIBentrySTDinterwordspacing

\bibitem{DBLP:journals/ijmmsc/BaeAB04}
\BIBentryALTinterwordspacing
M.~M. Bae, B.~F. AlBdaiwi, and B.~Bose, ``Edge-disjoint hamiltonian cycles in
  two-dimensional torus,'' \emph{Int. J. Math. Math. Sci.}, vol. 2004, no.~25,
  pp. 1299--1308, 2004. [Online]. Available:
  \url{https://doi.org/10.1155/S0161171204307325}
\BIBentrySTDinterwordspacing

\bibitem{8868152}
M.~Cho, U.~Finkler, M.~Serrano, D.~Kung, and H.~Hunter, ``Blueconnect:
  Decomposing all-reduce for deep learning on heterogeneous network
  hierarchy,'' \emph{IBM Journal of Research and Development}, vol.~63, no.~6,
  pp. 1:1--1:11, 2019.

\bibitem{thakur}
\BIBentryALTinterwordspacing
R.~Thakur, R.~Rabenseifner, and W.~Gropp, ``Optimization of collective
  communication operations in mpich,'' \emph{Int. J. High Perform. Comput.
  Appl.}, vol.~19, no.~1, p. 4966, feb 2005. [Online]. Available:
  \url{https://doi.org/10.1177/1094342005051521}
\BIBentrySTDinterwordspacing

\bibitem{nbc}
T.~Hoefler, A.~Lumsdaine, and W.~Rehm, ``{Implementation and Performance
  Analysis of Non-Blocking Collective Operations for MPI},'' in
  \emph{Proceedings of the 2007 International Conference on High Performance
  Computing, Networking, Storage and Analysis, SC07}.\hskip 1em plus 0.5em
  minus 0.4em\relax IEEE Computer Society/ACM, Nov. 2007.

\bibitem{isca_overlap}
\BIBentryALTinterwordspacing
S.~Rashidi, M.~Denton, S.~Sridharan, S.~Srinivasan, A.~Suresh, J.~Nie, and
  T.~Krishna, \emph{Enabling Compute-Communication Overlap in Distributed Deep
  Learning Training Platforms}.\hskip 1em plus 0.5em minus 0.4em\relax IEEE
  Press, 2021, p. 540–553. [Online]. Available:
  \url{https://doi.org/10.1109/ISCA52012.2021.00049}
\BIBentrySTDinterwordspacing

\bibitem{nik_alu}
N.~Dryden, N.~Maruyama, T.~Moon, T.~Benson, A.~Yoo, M.~Snir, and B.~Van~Essen,
  ``Aluminum: An asynchronous, gpu-aware communication library optimized for
  large-scale training of deep neural networks on hpc systems,'' in \emph{2018
  IEEE/ACM Machine Learning in HPC Environments (MLHPC)}, 2018, pp. 1--13.

\bibitem{you2017large}
Y.~You, I.~Gitman, and B.~Ginsburg, ``Large batch training of convolutional
  networks,'' 2017.

\bibitem{mathuriya2018cosmoflow}
A.~Mathuriya, D.~Bard, P.~Mendygral, L.~Meadows, J.~Arnemann, L.~Shao, S.~He,
  T.~Karna, D.~Moise, S.~J. Pennycook, K.~Maschoff, J.~Sewall, N.~Kumar, S.~Ho,
  M.~Ringenburg, Prabhat, and V.~Lee, ``Cosmoflow: Using deep learning to learn
  the universe at scale,'' 2018.

\bibitem{yang2020training}
J.~A. Yang, J.~Park, S.~Sridharan, and P.~T.~P. Tang, ``Training deep learning
  recommendation model with quantized collective communications,'' in
  \emph{Conference on Knowledge Discovery and Data Mining (KDD)}, 2020.

\bibitem{naumov2019deep}
M.~Naumov, D.~Mudigere, H.-J.~M. Shi, J.~Huang, N.~Sundaraman, J.~Park,
  X.~Wang, U.~Gupta, C.-J. Wu, A.~G. Azzolini \emph{et~al.}, ``Deep learning
  recommendation model for personalization and recommendation systems,''
  \emph{arXiv preprint arXiv:1906.00091}, 2019.

\bibitem{gupta2020architectural}
U.~Gupta, C.-J. Wu, X.~Wang, M.~Naumov, B.~Reagen, D.~Brooks, B.~Cottel,
  K.~Hazelwood, M.~Hempstead, B.~Jia \emph{et~al.}, ``The architectural
  implications of facebook's dnn-based personalized recommendation,'' in
  \emph{2020 IEEE International Symposium on High Performance Computer
  Architecture (HPCA)}.\hskip 1em plus 0.5em minus 0.4em\relax IEEE, 2020, pp.
  488--501.

\bibitem{mudigere2021software}
D.~Mudigere, Y.~Hao, J.~Huang, Z.~Jia, A.~Tulloch, S.~Sridharan, X.~Liu,
  M.~Ozdal, J.~Nie, J.~Park \emph{et~al.}, ``Software-hardware co-design for
  fast and scalable training of deep learning recommendation models,''
  \emph{arXiv preprint arXiv:2104.05158}, 2021.

\bibitem{brown2020language}
T.~B. Brown, B.~Mann, N.~Ryder, M.~Subbiah, J.~Kaplan, P.~Dhariwal,
  A.~Neelakantan, P.~Shyam, G.~Sastry, A.~Askell, S.~Agarwal, A.~Herbert-Voss,
  G.~Krueger, T.~Henighan, R.~Child, A.~Ramesh, D.~M. Ziegler, J.~Wu,
  C.~Winter, C.~Hesse, M.~Chen, E.~Sigler, M.~Litwin, S.~Gray, B.~Chess,
  J.~Clark, C.~Berner, S.~McCandlish, A.~Radford, I.~Sutskever, and D.~Amodei,
  ``Language models are few-shot learners,'' 2020.

\bibitem{shoeybi2020megatronlm}
M.~Shoeybi, M.~Patwary, R.~Puri, P.~LeGresley, J.~Casper, and B.~Catanzaro,
  ``Megatron-lm: Training multi-billion parameter language models using model
  parallelism,'' 2020.

\bibitem{gshard}
\BIBentryALTinterwordspacing
D.~Lepikhin, H.~Lee, Y.~Xu, D.~Chen, O.~Firat, Y.~Huang, M.~Krikun, N.~Shazeer,
  and Z.~Chen, ``Gshard: Scaling giant models with conditional computation and
  automatic sharding,'' 2020. [Online]. Available:
  \url{https://arxiv.org/abs/2006.16668}
\BIBentrySTDinterwordspacing

\bibitem{ibm-percs-network}
B.~Arimilli, R.~Arimilli, V.~Chung, S.~Clark, W.~Denzel, B.~Drerup, T.~Hoefler,
  J.~Joyner, J.~Lewis, J.~Li, N.~Ni, and R.~Rajamony, ``{The PERCS
  High-Performance Interconnect},'' in \emph{Proceedings of 18th Symposium on
  High-Performance Interconnects (Hot Interconnects 2010)}.\hskip 1em plus
  0.5em minus 0.4em\relax IEEE, Aug. 2010.

\bibitem{flatfly}
\BIBentryALTinterwordspacing
J.~Kim, W.~J. Dally, and D.~Abts, ``Flattened butterfly: A cost-efficient
  topology for high-radix networks,'' in \emph{Proceedings of the 34th Annual
  International Symposium on Computer Architecture}, ser. ISCA '07.\hskip 1em
  plus 0.5em minus 0.4em\relax New York, NY, USA: Association for Computing
  Machinery, 2007, p. 126–137. [Online]. Available:
  \url{https://doi.org/10.1145/1250662.1250679}
\BIBentrySTDinterwordspacing

\bibitem{hyperx}
J.~H. Ahn, N.~Binkert, A.~Davis, M.~McLaren, and R.~S. Schreiber, ``Hyperx:
  topology, routing, and packaging of efficient large-scale networks,'' in
  \emph{Proceedings of the Conference on High Performance Computing Networking,
  Storage and Analysis}, 2009, pp. 1--11.

\bibitem{keithcc}
\BIBentryALTinterwordspacing
K.~D. Underwood and E.~Borch, ``Exploiting communication and packaging locality
  for cost-effective large scale networks,'' in \emph{Proceedings of the 26th
  ACM International Conference on Supercomputing}, ser. ICS '12.\hskip 1em plus
  0.5em minus 0.4em\relax New York, NY, USA: Association for Computing
  Machinery, 2012, p. 291300. [Online]. Available:
  \url{https://doi.org/10.1145/2304576.2304616}
\BIBentrySTDinterwordspacing

\bibitem{BW}
B.~Bode, M.~Butler, T.~Dunning, T.~Hoefler, W.~Kramer, W.~Gropp, and
  H.~Wen-Mei, \emph{\BIBforeignlanguage{English (US)}{The blue waters
  super-system for super-science}}.\hskip 1em plus 0.5em minus 0.4em\relax CRC
  Press, Jan. 2013, pp. 339--366.

\bibitem{BGL}
N.~Adiga, G.~Almasi, G.~Almasi, Y.~Aridor, R.~Barik, D.~Beece, R.~Bellofatto,
  G.~Bhanot, R.~Bickford, M.~Blumrich, A.~Bright, J.~Brunheroto, C.~Cascaval,
  J.~Castanos, W.~Chan, L.~Ceze, P.~Coteus, S.~Chatterjee, D.~Chen, G.~Chiu,
  T.~Cipolla, P.~Crumley, K.~Desai, A.~Deutsch, T.~Domany, M.~Dombrowa,
  W.~Donath, M.~Eleftheriou, C.~Erway, J.~Esch, B.~Fitch, J.~Gagliano, A.~Gara,
  R.~Garg, R.~Germain, M.~Giampapa, B.~Gopalsamy, J.~Gunnels, M.~Gupta,
  F.~Gustavson, S.~Hall, R.~Haring, D.~Heidel, P.~Heidelberger, L.~Herger,
  D.~Hoenicke, R.~Jackson, T.~Jamal-Eddine, G.~Kopcsay, E.~Krevat, M.~Kurhekar,
  A.~Lanzetta, D.~Lieber, L.~Liu, M.~Lu, M.~Mendell, A.~Misra, Y.~Moatti,
  L.~Mok, J.~Moreira, B.~Nathanson, M.~Newton, M.~Ohmacht, A.~Oliner,
  V.~Pandit, R.~Pudota, R.~Rand, R.~Regan, B.~Rubin, A.~Ruehli, S.~Rus,
  R.~Sahoo, A.~Sanomiya, E.~Schenfeld, M.~Sharma, E.~Shmueli, S.~Singh,
  P.~Song, V.~Srinivasan, B.~Steinmacher-Burow, K.~Strauss, C.~Surovic,
  R.~Swetz, T.~Takken, R.~Tremaine, M.~Tsao, A.~Umamaheshwaran, P.~Verma,
  P.~Vranas, T.~Ward, M.~Wazlowski, W.~Barrett, C.~Engel, B.~Drehmel,
  B.~Hilgart, D.~Hill, F.~Kasemkhani, D.~Krolak, C.~Li, T.~Liebsch,
  J.~Marcella, A.~Muff, A.~Okomo, M.~Rouse, A.~Schram, M.~Tubbs, G.~Ulsh,
  C.~Wait, J.~Wittrup, M.~Bae, K.~Dockser, L.~Kissel, M.~Seager, J.~Vetter, and
  K.~Yates, ``An overview of the bluegene/l supercomputer,'' in \emph{SC '02:
  Proceedings of the 2002 ACM/IEEE Conference on Supercomputing}, 2002, pp.
  60--60.

\bibitem{tpusc}
\BIBentryALTinterwordspacing
N.~P. Jouppi, D.~H. Yoon, G.~Kurian, S.~Li, N.~Patil, J.~Laudon, C.~Young, and
  D.~Patterson, ``A domain-specific supercomputer for training deep neural
  networks,'' \emph{Commun. ACM}, vol.~63, no.~7, p. 67–78, jun 2020.
  [Online]. Available: \url{https://doi.org/10.1145/3360307}
\BIBentrySTDinterwordspacing

\bibitem{fugakunet}
{Yuichiro Ajima}, ``{High-dimensional Interconnect Technology for the K
  Computer and the Supercomputer Fugaku},'' Tech. Rep., 06 2019, fujitsu
  Technical Review.

\bibitem{9499662}
T.~T. Nguyen and M.~Wahib, ``An allreduce algorithm and network co-design for
  large-scale training of distributed deep learning,'' in \emph{2021 IEEE/ACM
  21st International Symposium on Cluster, Cloud and Internet Computing
  (CCGrid)}, 2021, pp. 396--405.

\bibitem{BML}
S.~Wang, D.~Li, Y.~Cheng, J.~Geng, Y.~Wang, S.~Wang, S.~Xia, and J.~Wu, ``A
  scalable, high-performance, and fault-tolerant network architecture for
  distributed machine learning,'' \emph{IEEE/ACM Transactions on Networking},
  vol.~28, no.~4, pp. 1752--1764, 2020.

\bibitem{haner-sim}
\BIBentryALTinterwordspacing
T.~H\"{a}ner and D.~S. Steiger, ``0.5 petabyte simulation of a 45-qubit quantum
  circuit,'' in \emph{Proceedings of the International Conference for High
  Performance Computing, Networking, Storage and Analysis}, ser. SC '17.\hskip
  1em plus 0.5em minus 0.4em\relax New York, NY, USA: Association for Computing
  Machinery, 2017. [Online]. Available:
  \url{https://doi.org/10.1145/3126908.3126947}
\BIBentrySTDinterwordspacing

\bibitem{greg-lu}
\BIBentryALTinterwordspacing
G.~Kwasniewski, M.~Kabic, T.~Ben-Nun, A.~N. Ziogas, J.~E. Saethre, A.~Gaillard,
  T.~Schneider, M.~Besta, A.~Kozhevnikov, J.~VandeVondele, and T.~Hoefler, ``On
  the parallel i/o optimality of linear algebra kernels: Near-optimal matrix
  factorizations,'' in \emph{Proceedings of the International Conference for
  High Performance Computing, Networking, Storage and Analysis}, ser. SC
  '21.\hskip 1em plus 0.5em minus 0.4em\relax New York, NY, USA: Association
  for Computing Machinery, 2021. [Online]. Available:
  \url{https://doi.org/10.1145/3458817.3476167}
\BIBentrySTDinterwordspacing

\bibitem{sharp}
R.~L. Graham, D.~Bureddy, P.~Lui, H.~Rosenstock, G.~Shainer, G.~Bloch,
  D.~Goldenerg, M.~Dubman, S.~Kotchubievsky, V.~Koushnir, L.~Levi, A.~Margolin,
  T.~Ronen, A.~Shpiner, O.~Wertheim, and E.~Zahavi, ``{Scalable Hierarchical
  Aggregation Protocol (SHArP): A Hardware Architecture for Efficient Data
  Reduction},'' in \emph{Proceedings of COM-HPC 2016: 1st Workshop on
  Optimization of Communication in HPC Runtime Systems - Held in conjunction
  with SC 2016: The International Conference for High Performance Computing,
  Networking, Storage and Analysis}.\hskip 1em plus 0.5em minus 0.4em\relax
  Institute of Electrical and Electronics Engineers Inc., jan 2017, pp. 1--10.

\bibitem{sharp2}
\BIBentryALTinterwordspacing
R.~L. Graham, L.~Levi, D.~Burredy, G.~Bloch, G.~Shainer, D.~Cho, G.~Elias,
  D.~Klein, J.~Ladd, O.~Maor, A.~Marelli, V.~Petrov, E.~Romlet, Y.~Qin, and
  I.~Zemah, ``{Scalable Hierarchical Aggregation and Reduction Protocol
  (SHARP)TM Streaming-Aggregation Hardware Design and Evaluation},'' in
  \emph{Lecture Notes in Computer Science (including subseries Lecture Notes in
  Artificial Intelligence and Lecture Notes in Bioinformatics)}, vol. 12151
  LNCS.\hskip 1em plus 0.5em minus 0.4em\relax Springer, jun 2020, pp. 41--59.
  [Online]. Available: \url{https://doi.org/10.1007/978-3-030-50743-5_3}
\BIBentrySTDinterwordspacing

\bibitem{flare}
D.~De~Sensi, S.~Di~Girolamo, S.~Ashkboos, S.~Li, and T.~Hoefler, ``Flare:
  Flexible in-network allreduce,'' in \emph{Proceedings of the International
  Conference for High Performance Computing, Networking, Storage and Analysis},
  ser. SC '21, 2021.

\bibitem{nvidiaallreduce}
B.~Klenk, N.~Jiang, G.~Thorson, and L.~Dennison, ``{An In-Network Architecture
  for Accelerating Shared-Memory Multiprocessor Collectives},'' 2020.

\bibitem{zero}
\BIBentryALTinterwordspacing
S.~Rajbhandari, J.~Rasley, O.~Ruwase, and Y.~He, ``Zero: Memory optimizations
  toward training trillion parameter models,'' 2019. [Online]. Available:
  \url{https://arxiv.org/abs/1910.02054}
\BIBentrySTDinterwordspacing

\bibitem{flexflow}
\BIBentryALTinterwordspacing
Z.~Jia, M.~Zaharia, and A.~Aiken, ``Beyond data and model parallelism for deep
  neural networks,'' 2018. [Online]. Available:
  \url{https://arxiv.org/abs/1807.05358}
\BIBentrySTDinterwordspacing

\bibitem{hxnet}
T.~Hoefler, M.~C. Heddes, and J.~R. Belk, ``Distributed processing
  architecture,'' U.S. Patent US11\,076\,210B1, Jul., 2021.

\bibitem{cglass}
T.~Hoefler, M.~C. Heddes, D.~Goel, and J.~R. Belk, ``Distributed processing
  architecture,'' U.S. Patent US20\,210\,209\,460A1, Jul., 2021.

\bibitem{wagma}
S.~Li, T.~Ben-Nun, G.~Nadiradze, S.~D. Girolamo, N.~Dryden, D.~Alistarh, and
  T.~Hoefler, ``{Breaking (Global) Barriers in Parallel Stochastic Optimization
  with Wait-Avoiding Group Averaging},'' \emph{IEEE Transactions on Parallel
  and Distributed Systems (TPDS)}, vol.~32, no.~7, pp. 1725--1739, 2021.

\bibitem{shigang-ppopp}
S.~Li and T.~Hoefler, ``{Near-Optimal Sparse Allreduce for Distributed Deep
  Learning},'' in \emph{Proceedings of the 27th ACM SIGPLAN Symposium on
  Principles and Practice of Parallel Programming}, Apr. 2022.

\bibitem{cosma}
G.~Kwasniewski, M.~Kabić, M.~Besta, J.~VandeVondele, R.~Solcà, and
  T.~Hoefler, ``{Red-Blue Pebbling Revisited: Near Optimal Parallel
  Matrix-Matrix Multiplication},'' in \emph{Proceedings of the International
  Conference for High Performance Computing, Networking, Storage and Analysis
  (SC19)}, Nov. 2019.

\bibitem{dryden2019improving}
N.~Dryden, N.~Maruyama, T.~Benson, T.~Moon, M.~Snir, and B.~V. Essen,
  ``Improving strong-scaling of cnn training by exploiting finer-grained
  parallelism,'' 2019.

\bibitem{vaswani2017attention}
A.~Vaswani, N.~Shazeer, N.~Parmar, J.~Uszkoreit, L.~Jones, A.~N. Gomez,
  L.~Kaiser, and I.~Polosukhin, ``Attention is all you need,'' 2017.

\bibitem{cscs}
``{Swiss National Supercomputing Centre - CSCS},'' \url{https://www.cscs.ch/},
  accessed: 30/03/2022.

\end{thebibliography}
	\appendix

	\section{Deep learning communication patterns}
	
	Section~\ref{sec:comm} provides a coarse overview of communications in deep learning. 
	Techniques for distributed and parallel deep learning are a major target for innovation and thus numerous approaches exist. Here, we provide some more background on various aspects that would overload the main text but show how most workloads can be mapped into the ring formulation today.
	
	Here, we focus on a brief overview of methods that lead to different communication requirements instead of an exhaustive list of all algorithms.
	
	\subsection{Data parallelism}\label{app:datapar}

	\paragraph{Local SGD with weight averaging} only communicate every $k$ iterations to reduce the communication complexity~\cite{localsgd1,stich2019local,gorbunov2020local,khaled2020tighter}. The communication sums all weights across all data-parallel processes and has thus the same communication pattern and volume as gradient summation. Some schemes perform only reductions on subgroups of the processes in order to reduce the communication volume further~\cite{wagma}.
	
	\paragraph{Gradient sparsification schemes} reduce the size of the gradient data, typically by sending only the largest gradients while accumulating the remaining ones into local buffers until they are large enough~\cite{dryden-quant,sparcml,NEURIPS2018_3328bdf9,NEURIPS2018_31445061}. This scheme is often combined with full communication steps in regular intervals. Reducing sparse gradients usually causes the vector to grow at each step due to fill-in. However, algorithms can be adapted to minimize the additional communication volume~\cite{shigang-ppopp}.
	
	\paragraph{Asynchronous SGD} performs weight updates asynchronously. This does not change the communication pattern in general but updates model copies independently~\cite{45187}. 
	
	\subsection{Operator parallelism}\label{app:oppar}
	
	Here, we discuss four typical operators: fully-connected (FC), convolutional (CV), transformer (TF), and element-wise (EL) operators. 
	It is easy to see that EL operators require no communication as they are independent, we will briefly discuss all others in the following.
	
	\paragraph{Fully-connected} FC operators are usually implemented by matrix multiplication on batches of examples, which may degenerate to matrix-vector multiplications for single examples. 
	Operator parallelism splits the activation matrix into $O$ pieces and performs a distributed matrix multiplication. The weight matrix can either be replicated at each process or also be distributed. 
	If it is replicated, then the forward pass does not require any communication but the backward pass needs to synchronize the weight update similar to data-parallelism using a reduction. 
	If the weight matrix is also distributed, then a full parallel matrix multiplication is performed. Communication-minimizing variants combine broadcast and distributed reductions~\cite{cosma}.
	
	\paragraph{Convolutional} CV operators are defined on a spatial domain and their parallelization follows a nearest-neighbor scheme. The simplest distribution splits the activations in one dimension, which leads to a nearest-neighbor communication scheme. 
	This scheme is often used in scientific computing workloads for the processing of high-resolution data~\cite{mathuriya2018cosmoflow,dryden2019improving}.
	
	\paragraph{Transformer} TF layers are among the largest layers used in practical workloads. For example, GPT-3~\cite{brown2020language} has 96 layers with 1.82 billion parameters each, resulting in 3.65GB of weight-storage (16-bit floats). 
	TF layers contain multiple operations (e.g., matmul, softmax, add)~\cite{vaswani2017attention}. They can be distributed with the Megatron-LM~\cite{shoeybi2020megatronlm} scheme that splits the inputs and requires one broadcast and two reductions in both the forward and backward pass.
	
	\paragraph{Summary} All operator-parallel schemes can be expressed efficiently with nearest-neighbor communication or broadcast and (all)reduce. 
	
	\subsection{Example Topologies}\label{app:topos}
	
	\subsubsection{Small cluster}
	
	We start with a small cluster design connecting 1,024 accelerators. We compare three different fat trees: nonblocking, 50\% tapered, and 75\% tapered, a full bandwidth Dragonfly, and Hx2Mesh as well as Hx4Mesh topologies. We also add a switchless 2D torus for comparison.
	
	\paragraph{Nonblocking fat tree}
	We use a two-level fat tree with 32 switches in the first and 16 switches in the second layer to reach 1,024 accelerators. This topology has 1,024 DAC cables connecting the accelerators and 1,024 AoC cables between the switches. 
	We use 16 planes, one for each network port per accelerator. 
	The total number of switches for 1,024 accelerators is (32+16)*16=768, AoC cables = DAC cables = 16*1,024. 
	
	\paragraph{Tapered fat trees}
	We use two level fat trees and implement tapering at the first level. For 50\%, we use 25 switches in the first layer, each of which routes 42 ports to accelerators and 22 ports to the second layer with 9 switches. 
	A total of 25*42=1,050 DAC cables connect to accelerators and 25*22=550 AoC cables connect to other switches. 
	The total number of switches for 1,050 accelerators is (25+9)*16=544. 
	
	For 75\% tapering, we use 21 switches in the first level, each with 51 ports to accelerators and 13 ports to the second layer with 5 switches.
	A total of 21*51=1,071 DAC cables connect to accelerators and 21*13=273 AoC cables connect to other switches. 
	The total number of switches for 1,071 accelerators is (21+5)*16=416. 
	
	\paragraph{Dragonfly}
	We assume a canonical Dragonfly configuration for 64 port switches ($a=2p=2h$~\cite{DF}): a group size of $a=16$ routers, $p=8$ terminals per router, and $h=8$ connections to other groups. This leads to a \emph{virtual switch} with 31 ports and we map two of those into a physical switch with 64 ports.
	Each group has 128 terminals and 8 physical switches (16 virtual). We use 8 groups to connect 1,024 endpoints. 
	Each group has 16 virtual switches with 8 AoC cables each connecting to other groups. This makes 8 groups*16 switches/group*8/2 cables per switch=512 AoC cables total. 
	Each group has 16 virtual switches with 8 endpoints each and 15 connections to other virtual switches. One inter-switch connection is internal to the physical switches. Thus, we have a total of 2*14 DAC cables per physical switch to other switches and 2*8 connections to endpoints per physical switch. With a total of 64 physical switches, we have 64*(2*14/2+2*8)=1,920 DAC cables.

	\paragraph{HxMesh}
	
	An Hx1Mesh would use $x=y=32$ to connect $32*32=1,024$ terminals. Each board has two connections per dimension (one for each direction) such that a 64-port switch can connect one row or column of boards. 
	This makes a total of 32+32 switches per plane. Each dimension has a total of $2*32*32=64*32=2,048$ cables to endpoints, making a total of 2,048 DAC and 2,048 AoC cables per plane.
	With four planes there are a total of 256 switches.
	
	An Hx2Mesh would use $x=y=16$ to connect $2*2*16*16=1,024$ terminals. 
	Each board has four connections per dimension (two per direction and two per row/column) such that a 64-port switch can connect one row or column of boards. 
	This makes a total of 16+16=32 switches per plane. Each dimension has a total of $4*16*16=16*64=1024$ cables to endpoints making a total of 1,024 DAC and 1,024 AoC cables per plane. 
	With our four planes there are a total of 128 switches.
	
	An Hx4Mesh would use $x=y=8$ to connect $4^2*8^2=1,024$ terminals.
	Each board has eight connections per dimension such that a 64-port switch can connect one row or column of boards. 
	This makes a total of 8+8=16 switches per plane.  Each dimension has a total of $8*64=512$ cables to endpoints making a total of 512 DAC and 512 AoC cables per plane. 
	With our four planes there are a total of 64 switches.
	
	\paragraph{2D torus}
	The 2D torus would be 32x32. With 2x2 PCB boards, there would be 16x16 boards to connect. Each board has 4 DAC connections (2 cables) in each dimension, making a total of $2*4/2*16*16=1,024$ DAC cables total.
	
	\subsubsection{Large cluster}
	
	\paragraph{Fat tree} A nonblocking fat tree would need three levels to connect 16,384 nodes --- the first two levels have 512 switches each and the third level would have 256 switches. The endpoints are attached with 16,384 DAC cables and the switches are connected with $2*16384$ AoC cables. 
	There are 16 planes. 
	
	The 50\% and 75\% tapered trees would be constructed similarly to the scheme above - the lowest level would route only 22 and 13 ports per switch to the second level, respectively. This requires a total of 794 switches, 17,160 AoC, and 16,380 DAC cables for 50\% and 8,304 switches, 8372 AoC, and 16,422 DAC cables for 75\% tapering for each of the 16 planes. 
	
	\paragraph{Dragonfly}
	We use a configuration wit a=32, p=17, h=16 that supports up to 279,072 terminals. We use 30 groups to construct a network with 16,320 terminals. Each plane uses 30*32=960 switches, 30*(17*32+31*32/2)=31,200 DAC, and 30*16*32/2=7,680 AoC cables
	
	\paragraph{HxMesh}
	
	An Hx1Mesh would use $x=y=128$ to connect $128*128=16,384$ terminals. 
	Both dimensions require fat trees of size $256$ (two ports per direction). We construct such fat trees with eight switches in the first layer and four switches in the second, twelve total. The $x$ direction requires 2*128*12=3,072 switches, the $y$ direction is analogous.
	This makes a total of $128*2*12$=3,072 switches per plane. One dimension can connect to the fat trees with DAC, there are a total of $2*128*128=32,768$ such links. The second level of the fat tree has the same number of AoC links. The other dimension's fat tree has twice the number of AoC links. 
	Since each plane consumes four ports (E, W, S, N), there are four planes for a total of 24,576 switches.
	
	The Hx2Mesh is a relatively expensive version because each accelerator is connected to two trees per plane.

	An Hx2Mesh would use $x=y=64$ to connect $2*2*64*64=16,384$ terminals. 
	Both dimensions require fat trees of size $128$ (two ports per direction). We construct such fat trees with four switches in the first layer and two switches in the second, six total. The $x$ direction requires 2*64*6=768 switches, the $y$ direction is analogous.
	This makes a total of $2*64*2*6$=1,536 switches per plane. One dimension can connect to the fat trees with DAC, there are a total of $2*64*128=16,384$ such links. The second level of the fat tree has the same number of AoC links. The other dimension's fat tree has twice the number. 
	Since each plane consumes four ports (E, W, S, N), there are four planes for a total of 6,144 switches.
	
	The Hx2Mesh is a relatively expensive version because each accelerator is connected to two trees per plane. 
	
	We note that another configuration where 2-level fat trees of size 2,048 are used to connect blocks of 16 rows and columns can lead to a higher global bandwidth at the same cost.
	
	An Hx4Mesh would use $x=y=32$ to connect $4*4*32*32=16,384$ terminals.
	Both directions can be connected by a single switch with $4$ switches per board per direction. Each board column in x direction has 64 ports, there are 4 switch columns per board, and 32 boards. Thus, the x direction is connected with 32*4*64=8,192 DAC cables and the y direction with the same number of AoC cables. 
	This makes a total of $2*32*4$=256 switches per plane, and a total of 1,024 switches for all four planes.
	
	\subsection{Disjoint Pipelined Rings}\label{app:hamiltonian}
	As discussed in Section~\ref{sec:fastred}, the allreduce can use at the same time all the four network interfaces of the accelerator, if implemented through two bidirectional pipelined rings mapped onto edge disjoint Hamiltonian cycles. To compute the two disjoint Hamiltonian cycle we used an existing algorithm~\cite{DBLP:journals/ijmmsc/BaeAB04} that works for 2D torus\footnote{An HxMesh can be seen as a 2D torus with additional links between boards.} of size $r \times c$, where $r$ is the number of rows, and $c$ the number of columns, if and only if $r = c \cdot k \; k \geq 1$ and $GCD(r, c-1) = 1$. 
	
	\begin{figure}[htpb]
		\begin{center}
			\hspace*{-1em}
			\includegraphics[width=\columnwidth]{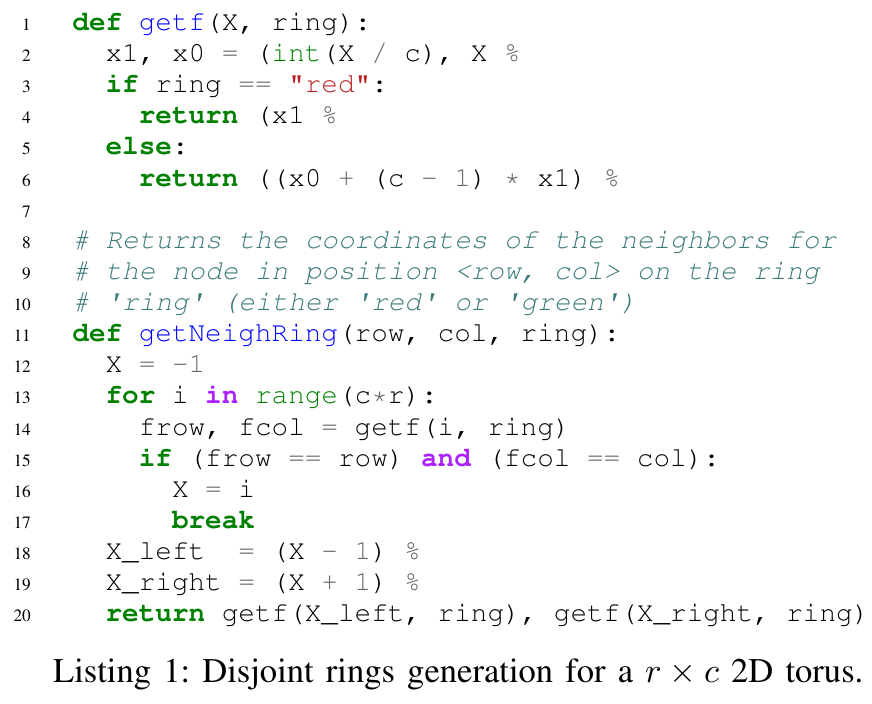}
		\end{center}
		\vspace{-1.0em}
		\label{fig:code}
	\end{figure}

	We describe the algorithm through Python code in Listing 1. Interest readers can find a formal definition with correctness proofs in the paper describing the algorithm~\cite{DBLP:journals/ijmmsc/BaeAB04}. The function \texttt{getf} of the algorithm (line 1) maps an identifier $X$ (between $0$ and the number of nodes in the torus) to a pair of coordinates on the torus. These identifiers are not assigned by a row major or column major order, and are uniquely identified by the \texttt{getf} function. Different mappings are used on the two different rings (that we denote with \textit{red} and \textit{green}). To find the neighbors of a specific node at coordinates \texttt{<row, col>}, the algorithm first finds the $X$ that is mapped to those coordinates (lines 13-17). Eventually, it computes the identifiers of the nodes on the left and on the right on the ring (lines 18-19), and their coordinates on the torus (lines 20-21).
	
\begin{figure}[htpb]
		\begin{center}
			\includegraphics[width=\columnwidth]{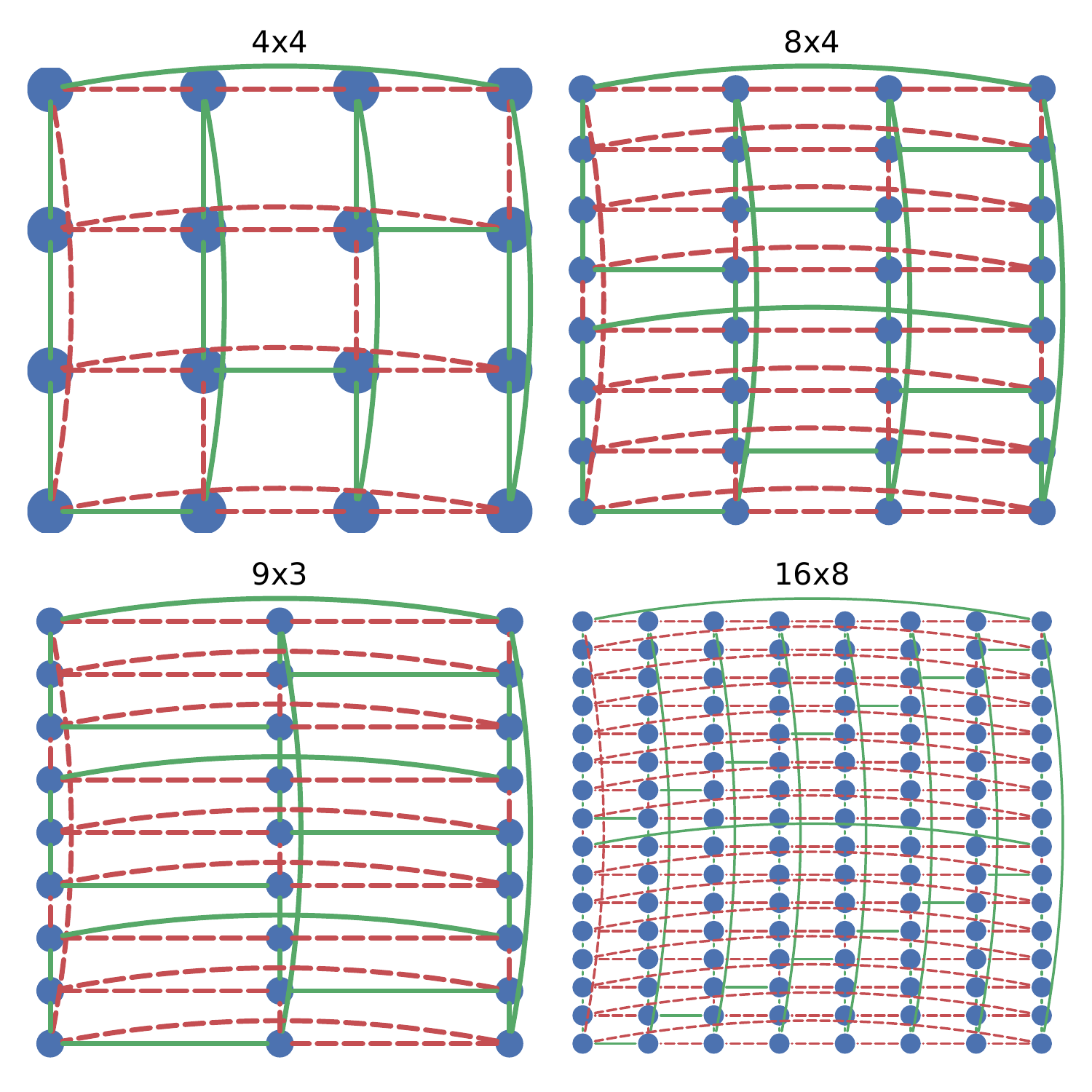}
		\end{center}
		\caption{Example of disjoint Hamiltonian cycles in 4x4, 8x4, 9x3, and 16x8 networks.}
		\label{fig:hamiltonian}
	\end{figure}

	We also report in Figure~\ref{fig:hamiltonian} the obtained disjoint rings for 2D torus of size $4x4$, $8x4$, $9x3$, and $16x8$. We observe that each node has two ports on each of the two rings, and that communications only occur between neighboring nodes. This allows each node to fully utilize all the available network interfaces.

	\subsection{Pricing Details}\label{app:pricing}
	
	\sloppy
	We sourced our pricing information on 3/25/22 on colfaxdirect.com for 100 Gb/s equipment. We assume that the relative prices are representative for future 400 Gb/s switches and cables. For the 64-port switches we chose Edgecore AS7816-64x 100G (https://www.colfaxdirect.com/store/pc/viewPrd.asp?idproduct =3592) at \$14,280 per piece. For 20m AoC cables, we picked Mellanox VCSEL-Based Active Fiber CableMellanox VCSEL-Based Active Fiber Cables at \$603 per piece  (https://www.colfaxdirect.com/store/pc/viewPrd.asp?idproduct =2892). For 5m DAC cables, we picked Mellanox Passive Copper Cable at \$272 per piece (https://www.colfaxdirect.com/store/pc/viewPrd.asp?idproduct =3546).
	
	\subsection{Simulation Setup}\label{app:sst}
	The Structural Simulation Toolkit (SST) version 11.1.0 has been used for the entirity of this paper. 
	The most important simulation parameters are shown in Table~\ref{tab:sst}. Notably, all simulations have been run using the eager protocol since rendezvous suffers from some critical bugs when using certain SST elements that are needed in our simulations.
	
		\begin{table}[ht!]
		\begin{center}
			\begin{tabular}{l|l}
				\textbf{Parameter}                                                          & \textbf{Value} \\ \hline
				Packet Size                                                                 & 8192 B         \\
				Flit Size                                                                   & 256 B          \\
				Buffer Sizes Per Port                                                       & 32 MB          \\ \hline
				Link Latency                                                                & 20 ns          \\
				Link Latency Hx Boards                                                      & 1 ns           \\
				In/Output Buffer Latency                                                    & 40 ns          \\ \hline
				\begin{tabular}[c]{@{}l@{}}Link BW \\ Simulating One Plane\end{tabular}     & 400 Gb/s       \\
				\begin{tabular}[c]{@{}l@{}}Link BW \\ Simulating All Planes\end{tabular}    & 1600 Gb/s      \\
				\begin{tabular}[c]{@{}l@{}}Cross Bar BW\\ Simulating One Plane\end{tabular} & 800 Gb/s       \\
				\begin{tabular}[c]{@{}l@{}}NIC BW\\ Simulating All Planes\end{tabular}      & 3200 Gb/s      \\
				\begin{tabular}[c]{@{}l@{}}Link BW \\ Simulating One Plane\end{tabular}     & 400 Gb/s       \\
				\begin{tabular}[c]{@{}l@{}}Link BW \\ Simulating All Planes\end{tabular}    & 1600 Gb/s      \\ \hline
				Number of Send Machines                                                     & 4             
			\end{tabular}
		\end{center}
		\caption{Main parameters used in SST when running the simulations.}
		\label{tab:sst}
	\end{table}
	
	The simulations have been run on various clusters with most of them running on:
	\begin{itemize}
		\item SlimFly cluster: An experimental internal cluster run in collaboration with CSCS~\cite{cscs}  composed of 192 nodes.
		\item Ault: A small cluster also run by CSCS. 8 nodes at a given time where used in this cluster.
	\end{itemize}
	
	All the benchmarks are completely automated and can be run from the appropriate folder where a readME file has been provided with the appropriate instructions.
	
	The best routing available in SST for each topology has been used when running the simulations (UGAL-L for Dragonfly, the routing described in Section~\ref{sec:routing} for HammingMesh and adaptive for the other topologies.)
	
	\subsection{Additional Results}\label{app:res}
	
\begin{figure}[ht!]
			\begin{center}
				\vspace{-1.3em}
				\includegraphics[width=\columnwidth]{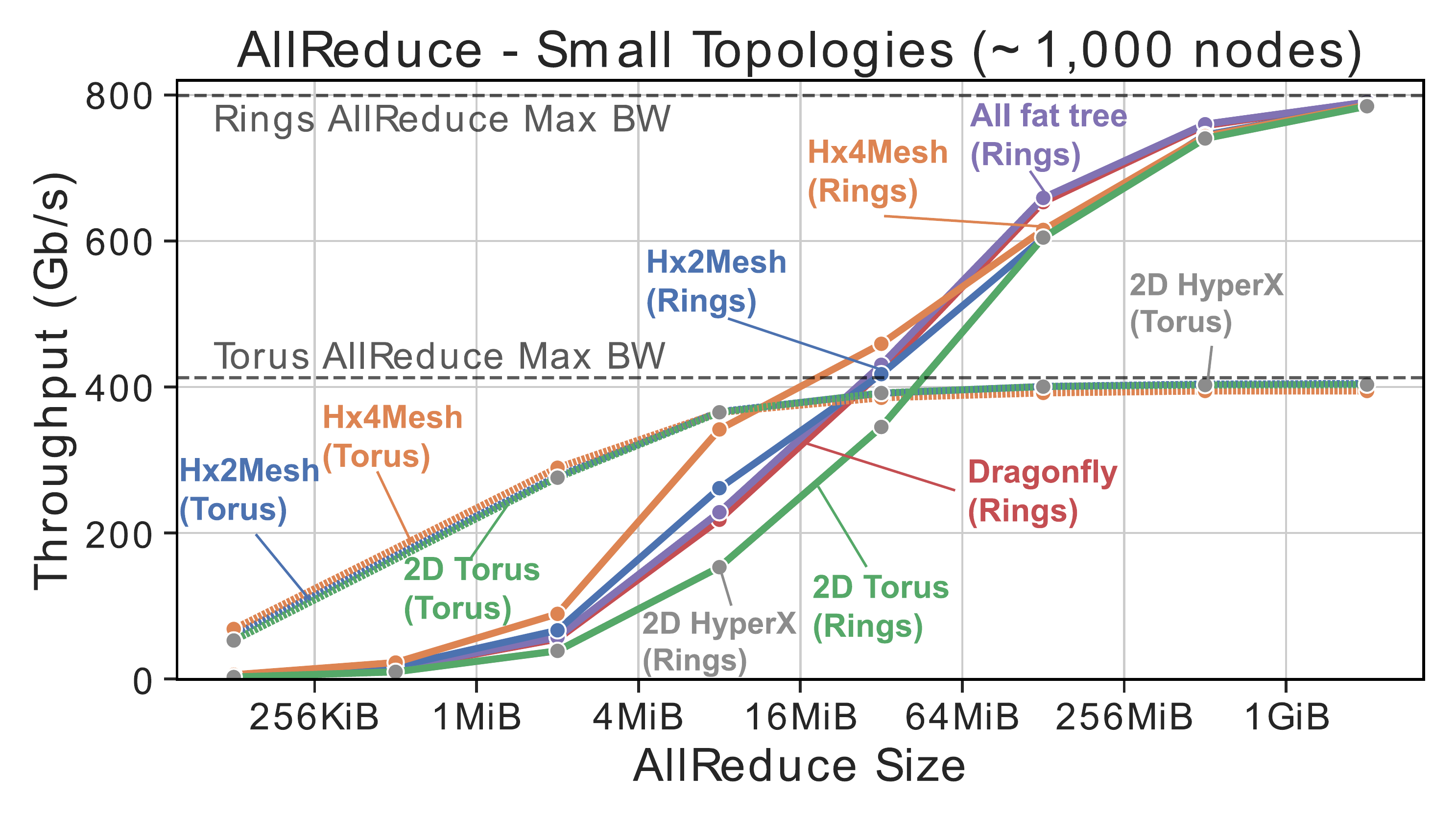}
			\end{center}
			\vspace{-1.3em}
			\caption{Global allreduce using different algorithms.}
			\vspace{-.99em}
			\label{fig:allredsmallapp}
	\end{figure}

		In this section we present some additional results. In particular we want to show the allreduce results also for the small topologies. We still use both the two bidirectional rings (``rings'') as well as the two-dimensional torus (``torus'') algorithm. As we can see the results are consistent with what we obtained with the large topologies in Section~\ref{sec:microbench}.
	Figure~\ref{fig:allredsmallapp} shows the achieved bandwidths.
	
\end{document}